\documentclass[sigconf]{acmart}
\renewcommand\footnotetextcopyrightpermission[1]{} 
\fancypagestyle{firstpage}{
	\fancyhf{}
	\setlength{\headheight}{50pt}
	
	\pagenumbering{arabic}
}  

\usepackage{xcolor}
\usepackage{subfig}
\usepackage{multirow}
\usepackage{fancyhdr}
\usepackage{amsmath}
\usepackage{xspace}
\usepackage{soul}
\usepackage{placeins}
\usepackage{array,booktabs,multirow}

\usepackage{subfig}
\usepackage{url}
\usepackage{etex}
\usepackage{etoolbox}
\usepackage{xspace}
\usepackage{amsmath}
\usepackage{xcolor}
\usepackage{nicefrac}
\usepackage{changepage}
\usepackage{array,booktabs,multirow}
\usepackage{afterpage}
\usepackage{pbox}
\RequirePackage{enumerate}
\RequirePackage{paralist}
\RequirePackage{enumitem}
\usepackage{paralist}
\usepackage{mdwlist}

\usepackage{hhline}
\usepackage{lipsum} 
\usepackage{adjustbox}
\usepackage{tabu}
\usepackage{makecell}
\usepackage{color, colortbl}
\usepackage{amsmath,amssymb,latexsym}

\usepackage{algorithm}
\usepackage[noend]{algpseudocode}

\usepackage{mathastext}

\usepackage{blindtext}
\usepackage[absolute]{textpos}

\usepackage{amsfonts}

\usepackage{pifont}

\makeatletter                                                                             
\newcommand{\thickhline}{%
    \noalign {\ifnum 0=`}\fi \hrule height 1.2pt                                          
    \futurelet \reserved@a \@xhline                                                       
}                                                                                         
\makeatletter                                                                             
\newcommand{\midhline}{%
    \noalign {\ifnum 0=`}\fi \hrule height 1pt                                            
    \futurelet \reserved@a \@xhline                                                       
}    

\usepackage{algorithm}
\usepackage[noend]{algpseudocode}

\floatevery{algorithm}{\setlength\hsize{3.7cm}}

\algnewcommand{\LineComment}[1]{\State \(\triangleright\) #1}     
\makeatletter
\renewcommand{\ALG@beginalgorithmic}{\small}
\makeatother                        
                                                                                          
\algblockdefx{FORP}{ENDFP}[1]%
{\textbf{for all }#1 \textbf{do in parallel}}%
{\textbf{end for}}

\algblockdefx{FORB}{ENDFB}[1]%
{\textbf{for }#1 \textbf{do in backward}}%
{\textbf{end for}}

\makeatletter                                                                             
\newcommand{\algorithmfootnote}[2][\footnotesize]{%
    \let\old@algocf@finish\@algocf@finish
    \def\@algocf@finish{\old@algocf@finish
        \leavevmode\rlap{\begin{minipage}{\linewidth}                                     
                #1#2                                                                      
        \end{minipage}}%
    }%

}             

\makeatletter  
\def\@eqnnum{{\normalfont\normalcolor[\theequation]}}  
\makeatother

\newcommand{\sect}[1]{Sec.~\ref{#1}\xspace}

\makeatletter
\renewcommand{\fnum@table}{Tab. \thetable}
\makeatother
\newcommand{\tab}[1]{Tab.~\ref{#1}\xspace}

\makeatletter
\renewcommand{\fnum@figure}{Fig. \thefigure}
\makeatother
\newcommand{\fig}[1]{Fig.~\ref{#1}\xspace}

\newcommand{\scratch}[1]{}

\makeatletter
 \def\SOUL@hlpreamble{%
 \setul{0ex}{2ex}
 \let\SOUL@stcolor\SOUL@hlcolor
 \SOUL@stpreamble
 }
\makeatother
\soulregister\cite7
\soulregister\ref7
\soulregister\pageref7
\soulregister\fig7
\soulregister\Sect7
\soulregister\Fig7
\soulregister\Tbl7
\soulregister\Equ7
\soulregister\Alg7
\soulregister\sect7
\soulregister\fig7
\soulregister\tbl7
\soulregister\equ7
\soulregister\alg7
\soulregister\gpusnap7

\newcolumntype{L}[1]{>{\raggedright\let\newline\\\arraybackslash\hspace{0pt}}m{#1}}
\newcolumntype{C}[1]{>{\centering\let\newline\\\arraybackslash\hspace{0pt}}m{#1}}
\newcolumntype{R}[1]{>{\raggedleft\let\newline\\\arraybackslash\hspace{0pt}}m{#1}}
\newcolumntype{M}[1]{>{\raggedright\arraybackslash}m{#1}}

\SetProtrusion{encoding={*},family={*},size={10,11,12}}
{-={1000,1000},
        ”={,1000},
        “={1000,},
        ‘={1000,},
        ’={,1000},
        :={,1000},
        ?={,1000},
        {,}={,1000},
        .={,1000},
        ;={,1000},
        )={,1000},
        ]={,1000},
        a={0,0},
        b={0,0},
        c={0,0},
        d={0,0},
        e={0,0},
        f={0,0},
        g={0,0},
        h={0,0},
        i={0,0},
        j={0,0},
        k={0,0},
        l={0,0},
        m={0,0},
        n={0,0},
        o={0,0},
        p={0,0},
        q={0,0},
        r={0,0},
        s={0,0},
        t={0,100},
        u={0,0},
        v={0,0},
        w={0,0},
        x={0,0},
        y={0,0},
        z={0,0},
        A={0,0},
        B={0,0},
        C={0,0},
        D={0,0},
        E={0,0},
        F={0,0},
        G={0,0},
        H={0,0},
        I={0,0},
        J={0,0},
        K={0,0},
        L={0,0},
        M={0,0},
        N={0,0},
        O={0,0},
        P={0,0},
        Q={0,0},
        R={0,0},
        S={0,0},
        T={0,0},
        U={0,0},
        V={0,0},
        W={0,0},
        X={0,0},
        Y={0,0},
        Z={0,0},
        0={0,0},
        1={0,0},
        2={0,0},
        3={0,0},
        4={0,0},
        5={0,0},
        6={0,0},
        7={0,0},
        8={0,0},
        9={0,0}}  

\usepackage[moderate,leading=normal,tracking=normal,wordspacing=normal,lists=normal,bibnotes=normal]{savetrees}

\setitemize{noitemsep,topsep=6pt,parsep=6pt,partopsep=0pt}

\newlength{\bibitemsep}
\newlength{\bibparskip}
\let\oldthebibliography\thebibliography
\renewcommand\thebibliography[1]{%
  \oldthebibliography{#1}%
  \setlength{\parskip}{\bibitemsep}%
  \setlength{\itemsep}{\bibparskip}%
}
\newcommand{\mnew}[1]{{\color{black}{#1}}}

\newcommand{\mdel}[2]{}
\newcommand{\FIXME}[1]{}

\title[Buddy Compression: Enabling Larger Memory for Deep Learning and HPC...]{Buddy Compression: Enabling Larger Memory for Deep Learning and HPC Workloads on GPUs}

 \author{Esha Choukse}
 \affiliation{%
   \institution{University of Texas at Austin}
 }
 \email{esha.choukse@utexas.edu}
 \author{Michael B. Sullivan}
 \affiliation{%
   \institution{NVIDIA}
 }
 \email{misullivan@nvidia.com}
 \author{Mike O'Connor}
 \affiliation{%
   \institution{NVIDIA}
 }
 \email{moconnor@nvidia.com}
 \author{Mattan Erez}
 \affiliation{%
   \institution{University of Texas at Austin}
 }
 \email{mattan.erez@utexas.edu}
 \author{Jeff Pool}
 \affiliation{%
   \institution{NVIDIA}
 }
 \email{jpool@nvidia.com}
 \author{David Nellans}
 \affiliation{%
   \institution{NVIDIA}
 }
 \email{dnellans@nvidia.com}
 \author{Stephen W. Keckler}
 \affiliation{%
   \institution{NVIDIA}
 }
 \email{skeckler@nvidia.com}

\renewcommand{\shortauthors}{E. Choukse et al.}

\begin{document}
\sloppy

\captionsetup{skip=3pt}
\captionsetup{font={small,bf}}


\begin{abstract}

GPUs offer orders-of-magnitude higher memory bandwidth than traditional
CPU-only systems. But, their memory capacity tends to be relatively small and
can not be increased by the user. This work proposes
Buddy Compression, a scheme to increase both the effective GPU memory capacity,
and bandwidth, while avoiding the downsides of conventional memory-expansion
techniques.
Buddy Compression splits each compressed
memory-entry between high-speed GPU memory and a slower-but-larger
disaggregated memory pool or host-CPU memory, such that highly-compressible memory-entries are accessed completely from GPU memory, while
incompressible entries source some of their data from off-GPU memory.
We show that Buddy Compression achieves an average compression ratio of $1.9$x for representative HPC applications and $1.5$x for deep learning workloads, with a performance within $2\%$ of an ideal system containing 100\% high speed and high capacity memory.
This makes Buddy Compression an ideal candidate for developers that require some additional memory capacity, and can tolerate minimal performance penalty.
\end{abstract}

\maketitle
\section{Introduction}

GPUs are widely used for many high-memory-footprint applications, including
those for High Performance Computing (HPC) and Deep Learning (DL).
HPC applications
like weather prediction and the modeling of fluid and molecular dynamics have
grown to require very large models~\cite{largehpc1,largehpc2}. DL networks are
also developing in a direction where either the model sizes are too big
to run on GPUs, or they are large enough such that the only a small batch
size can fit on the GPU, resulting in low utilization, and accuracy
issues~\cite{megdet2017,chen_modnn:_2018,BERT}. 
Today, applications with large memory footprints must:
\begin{inparaenum}[\itshape (i)]
\item scale out to many GPUs for capacity purposes (inefficient resource
utilization)~\cite{wang_supporting_2018,akiba_pfdet:_2018},
\item explicitly orchestrate data movement between the host CPU and the GPU to stay within device memory
limitations (adding algorithmic complexity)~\cite{rhu_vdnn:_2016,chen_modnn:_2018, rhu_compressing_2018},
or \item rely on off-GPU memory accesses or Unified Memory~\cite{harris_2017} to automatically oversubscribe
device memory (limiting performance)~\cite{tianhao_uvm,ammar_ahmad_awan_can_nodate}.
\end{inparaenum}
In this paper, we explore memory compression as a solution to
this challenge.

Main memory compression has been studied in detail for
CPUs~\cite{mxt2001,rmc2005,lcp2013,Buri,compresso2018}. However,
GPU-specific workloads and architectural details
pose very different trade-offs. For instance, the CPU solutions assume the
compressed pages to be of different sizes and that they can be re-allocated as the
compression ratios change~\cite{rmc2005,lcp2013,Buri,compresso2018}. Due to
the immense device memory bandwidth, relying on such on-the-fly page
re-allocations has a huge impact on throughput in GPUs~\cite{tianhao_uvm}. 

While domain-specific compression
~\cite{nystad_adaptive_2012,jain_gist:_2018} has been
explored to help with large workloads on GPUs,
hardware memory compression to increase memory capacity
for general purpose applications in GPUs
remains unexplored.

\begin{figure}
  \vspace{-0.5em}
  \centering
  \includegraphics[width=0.90\columnwidth]{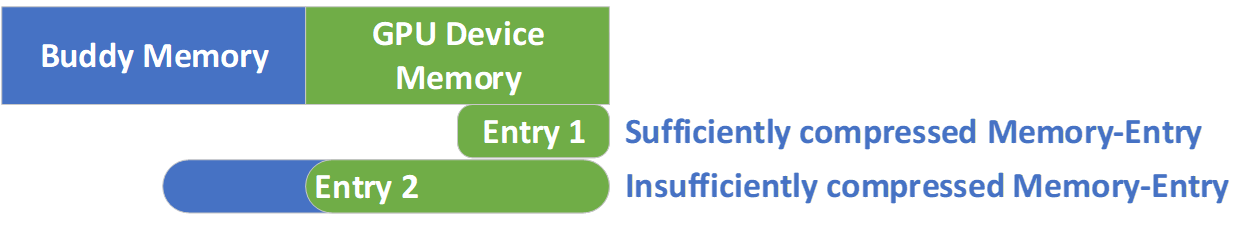}
  \caption{In Buddy Compression, if a memory-entry (128B) does not compress sufficiently, part of it is accessed from the buddy-memory.}
  \label{fig:buddy_basic}
\end{figure}

\begin{figure}
  \centering
  \includegraphics[width=0.9\columnwidth]{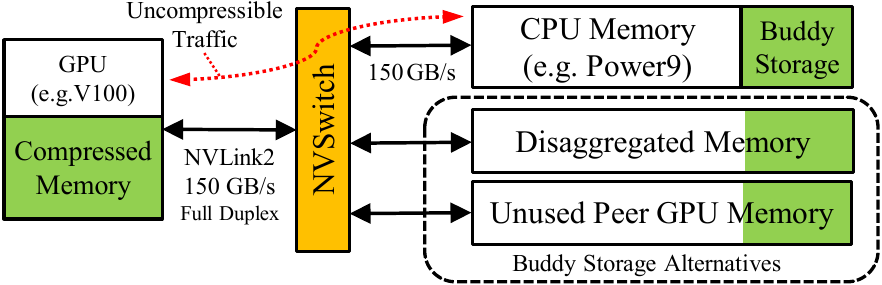}
  \caption{A target system for Buddy Compression. Any larger NVLink-connected memory is used as buddy storage. Overall organization is like NVIDIA DGX-2~\cite{dgx2}.}
  \label{fig:overview_fig}
  \vspace{-1.5em}
\end{figure}


\mnew{In Buddy Compression, we compress the data and divide the compressed memory
allocations between the GPU device memory and a larger-but-slower
buddy-memory connected with a high-bandwidth interconnect (Figure~\ref{fig:buddy_basic}).
If a memory-entry is sufficiently compressed, it is sourced completely from
device memory. If not, it is sourced from both device and buddy-memory.
This design requires no re-allocations within the device memory
as the data changes compressibility over time.
A high-bandwidth interconnect like NVLink~\cite{dgx2}, OpenCAPI~\cite{opencapi} or PCIe5.0~\cite{intelCXL} is the enabling feature for this design, since it ensures
low overhead accesses to the buddy-memory, as long as most of the data is in GPU device memory.
Any remote memory that is connected to the GPU with a high-bandwidth interconnect is suitable for
being used as buddy-memory (Figure~\ref{fig:overview_fig}).}
As we demonstrate in the rest of the paper, this design maintains a good compression ratio and high
performance, while avoiding the complexity and performance concerns of mapping CPU
memory compression approaches on GPUs. To summarize the research contributions of this work:
\begin{itemize}[leftmargin=*]
\item We provide an in-depth analysis of the data of representative GPU workloads and derive insights
for effective GPU compression.
\item We introduce the first design to use general-purpose compression to increase
the memory capacity of GPUs. Buddy Compression is unique,
since it does not require any additional data movement as the compressibility of the data changes.
\item We show that Buddy Compression \mnew{achieves 1.9x compression and} performs within 2\% of an ideal, high-memory-capacity GPU.
\item Finally, we present a case study on DL training to understand the benefits and trade-offs of using Buddy
Compression.
\end{itemize} 


\section{Overview}
\begin{table}[b]
\vspace{-0.5em}
\setlength{\tabcolsep}{0pt}
\centering
\caption{Details of the GPU Benchmarks Used}
\label{tab:config}
\footnotesize
\begin{tabular*}{\columnwidth}{@{\hspace{0.05\columnwidth}}L{0.25\columnwidth}@{}R{0.15\columnwidth}@{\hspace{0.05\columnwidth}}L{0.3\columnwidth}@{}R{0.15\columnwidth}@{\hspace{0.05\columnwidth}}}
\toprule
\multicolumn{2}{l}{\hspace{0.05\columnwidth}\textbf{HPC SpecAccel}}       & \multicolumn{2}{l}{\textbf{HPC FastForward}}  \\
351.palm                             & 2.89GB & FF\_HPGMG-FV    & 2.32GB        \\
352.ep                               & 2.75GB & FF\_Lulesh      & 1.59GB        \\
354.cg                               & 1.23GB & \multicolumn{2}{l}{\textbf{DL Training}}  \\ 
355.seismic                          & 2.83GB & BigLSTM			& 2.71GB		\\
356.sp                               & 2.83GB & AlexNet         & 8.85GB        \\
357.csp                              & 1.44GB  & Inception\_V2   & 3.21GB       \\
360.ilbdc                            & 1.94GB & SqueezeNetv1.1  & 2.03GB       \\
370.bt                               & 1.21MB  & VGG16           & 11.08GB        \\
                                     &        & ResNet50        & 4.50GB       \\
\bottomrule
\end{tabular*}
\vspace{-10pt}
\end{table}
\normalsize


\subsection{Target Workloads and Motivation}

\textbf{HPC Workloads}.  Several important HPC applications, ranging from fluid
dynamics to weather prediction, have found GPUs to be the accelerator of
choice. These models have outgrown GPU device memory~\cite{largehpc1,largehpc2}.  Today, scientists use
either Unified Memory or multiple GPUs to scale the models. Device memory
compression can be very useful for such cases. We use a subset of SpecAccel and
DOE FastForward benchmarks to represent these HPC applications. The subset is
chosen based on the confidence in the representativeness of the data values
used in the benchmarks.
All the discarded benchmarks seemed to have large portions
of their working sets be zero, thereby having extremely high compression
ratios.

\textbf{DL Workloads}. GPUs are currently the most popular choice for training
deep neural networks. As these networks grow deeper and wider, they require
more data and are inevitably hitting the memory-capacity wall, as we discuss in
detail in Section~\ref{sec:dl_stuff}.  While many domain-specific solutions
across the stack have been proposed to deal with this memory capacity challenge
in deep learning
training~\cite{dsg2018,dlcheckpoint2016,vdnn2016,lowprecision2018,jain_gist:_2018,wang_supporting_2018,ito_ooc_cudnn:_2017,rhu_compressing_2018},
our proposal requires no algorithm-level changes, and applies to other classes
of workloads as well.  We use a set of convolutional neural networks (CNNs) and
one recurrent neural network (RNN) to represent this class of workloads in our
evaluation.

\subsection{Related Work}
There are two approaches to tackle memory limitations: compression, and
domain-specific techniques.
The graphics pipeline of most GPUs includes texture memory that is lossily compressed offline
using tailored compression algorithms~\cite{nystad_adaptive_2012,nvidia_turing} in order to
reduce the footprint of these textures.
\mnew{The deep learning space has domain-specific solutions across the stack
~\cite{dlcheckpoint2016,jain_gist:_2018,vdnn2016}.
Buddy Compression is orthogonal to most of these proposals.
For instance, vDNN~\cite{vdnn2016}
proposes running large DL networks using manual offloading of data
layer-by-layer. However, there are still cases where it fails, due to the
inability to fit data required for just one layer~\cite{uvm_vdnn}. Buddy
compression can enable running larger networks with vDNN}.
To our knowledge, hardware compression is not currently used for general-purpose compute workloads on GPUs.

In CPUs, for decades, memory compression
in various forms has been proposed and used as a solution to the memory
capacity challenge. Most modern operating systems compress the swap
space to reduce paging to disk~\cite{zram}.
There have been numerous proposals
for hardware-accelerated main memory
compression~\cite{mxt2001,rmc2005,lcp2013,Buri,compresso2018}.

\subsection{Relevant Modern GPU Technology}

\textbf{Unified Memory (UM).} Unified Memory (UM), introduced in
CUDA 8 for Pascal-class GPUs~\cite{harris_2017}, allows sharing a single memory space
across a heterogeneous node. Non-local UM requests either remotely
access data through the GPU interconnect or result in transparent data
migration with the placement of any piece of data being determined by a variety
of heuristics~\cite{harris_2017,uvm_hpgmg}.
UM supports memory over-subscription, allowing UM-managed regions that are
larger than the GPU device memory to be accessed without explicit data
movement.
This has not been widely adopted, since applications with large hot working sets experience frequent
page faults and thrashing with Unified Memory, causing high performance
overheads~\cite{uvm_hpgmg,uvm_cuda_app,uvm_vdnn}. 
\mnew{Similar solutions are available for AMD and ARM GPUs using
Heterogeneous System Architecture (HSA) and CCIX~\cite{hsaAMD1,hsaAMD2},
and very recently, for Intel GPUs using Compute eXpress Link (CXL)~\cite{intelCXL}.}

\textbf{High Bandwidth Interconnects.}
In recent years, high bandwidth interconnects like
NVLink~\cite{tiffany_trader_tsubame3.0_nodate}, openCAPI~\cite{opencapi},
and NVLink2~\cite{dgx2} have been used to alleviate the communication bottleneck
in multi-GPU systems.
\mnew{Buddy Compression is made possible due to these high-bandwidth interconnects.}
NVLink2 provides 25GBps of full-duplex unidirectional bandwidth per \textit{brick}. Modern
compute-class V100 GPUs support six NVLink2 bricks per GPU, offering a
bidirectional bandwidth of up to 150GBps (full-duplex), much higher than the 16GBps
$\times16$ PCIe3.0 full-duplex connection. The NVIDIA DGX-2~\cite{dgx2}
workstation has sixteen V100 GPUs connected through an NVLink2 switch that
supports the full 150GBps traffic between any two GPUs. IBM Power9 CPUs also
support six NVLink2 bricks, allowing high-speed remote access to system
memory~\cite{ibmp9}.

\textbf{Buddy Compression Target System.}
Given the trends in modern
GPU nodes, the future-facing system we envision for Buddy Compression is shown
in Figure~\ref{fig:overview_fig}. It is composed of NVSwitch-connected multi-GPU
nodes with NVLink2-based access to a larger source of remote memory.
In currently available systems, this
remote memory could be the system memory of a Power9 CPU, or unused peer GPU
memory. While we know of no
current NVLink-connected disaggregated memory appliance, such a device is a
natural extension of the technology that is being explored for
servers~\cite{dragonDisaggregated, parthaDisaggregated}.
\mnew{THe high-bandwidth interconnect is what enables Buddy Compression.
So, as long as the remote
memory sources operate at the full NVLink2 bandwidth, Buddy Compression applies
equally well. Any high bandwidth interconnect can be used in place of NVLink2.}

\subsection{Compression Design Considerations}\label{sec:comp}
\begin{figure*}[h]
  \vspace{-1em}  
  \centering
  \includegraphics[width=0.7\textwidth]{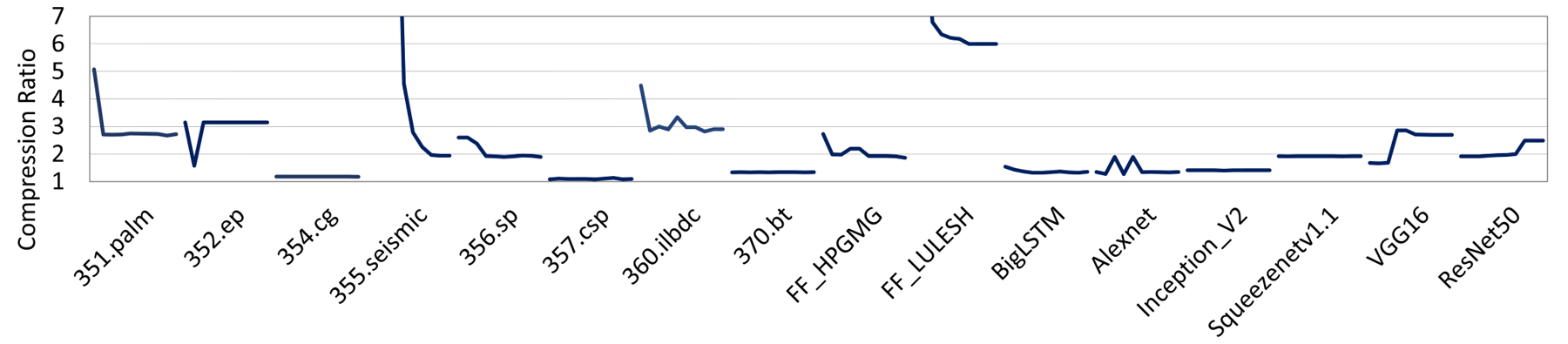}
    \caption{The average compression ratio of the allocated memory for the complete run of each benchmark. Ten equally distributed memory-snapshots are taken during the entire run of each benchmark, and the compression ratio calculated.}
  \label{fig:overtime}
  \vspace{-1em}
\end{figure*}
There are some important design
choices and challenges that need to be addressed in a compressed memory proposal. We
present these design points in brief.

{\bf Compression Algorithms.} A hardware memory compression algorithm should be
fast and require little energy, yet result in high compression rates. \textit{After comparing
several algorithms~\cite{bdi2012,fpc,fvc,cpack2010,bpc2016}, we choose Bit-Plane Compression
(BPC)~\cite{bpc2016} for Buddy Compression}. BPC has been shown to have high
compression ratios for GPU benchmarks when applied for DRAM bandwidth
compression.

{\bf Compression Granularity.} This is the unit of data that is compressed or
decompressed together. A higher
compression granularity requires less metadata, and generally results in higher
compression. On the other hand, a lower compression granularity does not
require as many read-modify-write operations.
Most CPU main memory compression work uses a
cache-block sized compression granularity to avoid these overheads.
\textit{We share this design decision,
and, following the results of microbenchmarks~\cite{jia_dissecting_2018}, use a 
128B memory-entry as the compression granularity for GPUs.}


{\bf Translation Metadata.} The compressed size of each 128B memory-entry
depends on its compressibility. This
requires additional translation metadata to access compressed data.
This metadata lookup generally
lies on the critical path for every access to the memory.
The layout of a compressed main-memory is somewhat similar in all previous work on main memory compression
in CPUs. There is some space dedicated for the compression metadata, and the
rest is arranged as variable-sized pages. Page size is decided by the
compressibility of the data within a page. 



{\bf Compressed Data Movement and Allocation.} As data is written back to a
compressed memory, value changes can lead to changes in compressibility. This
means that a memory-entry can grow or shrink over time, leading to additional
data movement~\cite{compresso2018}. Allocation granularity is closely related
to the data movement overhead.
For example, current systems allocate memory at page granularity.
Changes in the size of one memory-entry can lead to data movement within and across the pages.
However, if each memory-entry is separately allocated and translated,
its size changes do not affect other data.



\section{Buddy Compression}

\subsection{Compressibility of Workloads} To estimate the possible gains
from compression, it is imperative to first find how compressible the
high-footprint GPU workloads are. To this end, we take memory dumps of the
workloads running on a Tesla P100 GPU with an Intel Xeon E5 host. We intercept
each GPU \textit{malloc} and \textit{free} API call (including variants for
pinned and Unified Memory-managed memory) to dynamically track the current allocated
regions in device memory. We divide the entire runtime of the workload into
10 regions, and at kernel boundaries of each region, take a memory dump of the
allocated device memory.     



Figure~\ref{fig:overtime} shows the compression ratio of each benchmark using
BPC compression~\cite{bpc2016} over its entire run. Note that these compression
ratios are quite optimistic capacity compression, since they assume eight
different compressed memory-entry sizes are available ($0B$, $8B$, $16B$,
$32B$, $64B$, $80B$, $96B$, and $128B$) and assume no page-packing overheads.
That is, each memory-entry is individually compressed and allowed to occupy any
of the above mentioned sizes. On average, the geometric mean of compression
ratio for the HPC workloads is $2.51$ for the HPC benchmarks and $1.85$ for the
DL benchmarks. This is a higher average as compared to prior work on CPU
workloads~\cite{compresso2018}, and can be attributed to the higher
percentage of homogeneous data allocations (with a single uniform datatype).
Prior work has established that Bit-Plane Compression works well for homogeneous
data, and such homogeneity is prevalent in GPU workloads~\cite{bpc2016}.

{\bf Compressibility Changes.} As compared to previously studied
workloads~\cite{lcp2013,compresso2018,rmc2005}, the compressibility changes
over time more often in GPU benchmarks. The most extreme example is $355.seismic$, which begins
with many zero values but slowly asymptotes to a 2x compression ratio over its
execution. We also observe that although the overall compression ratio of the
DL workloads stays roughly constant, there are frequent compressibility changes
for individual memory entries. This is likely due to the fact that DL
frameworks perform their own asynchronous GPU memory allocation with software-managed 
memory pools, and may reuse the same memory location for a variety of purposes over program
execution~\cite{caffe}. In some cases, the data in a memory-entry can grow
more random over time, thereby decreasing its compressibility. This would
require more space to be allocated for the same memory-entry, causing a
\textit{memory-entry overflow}, and thereby additional data movement, as
discussed in Section~\ref{sec:comp}.



\subsection{Buddy Compression Overview}
\begin{figure}
  \centering
  \includegraphics[width=0.80\columnwidth]{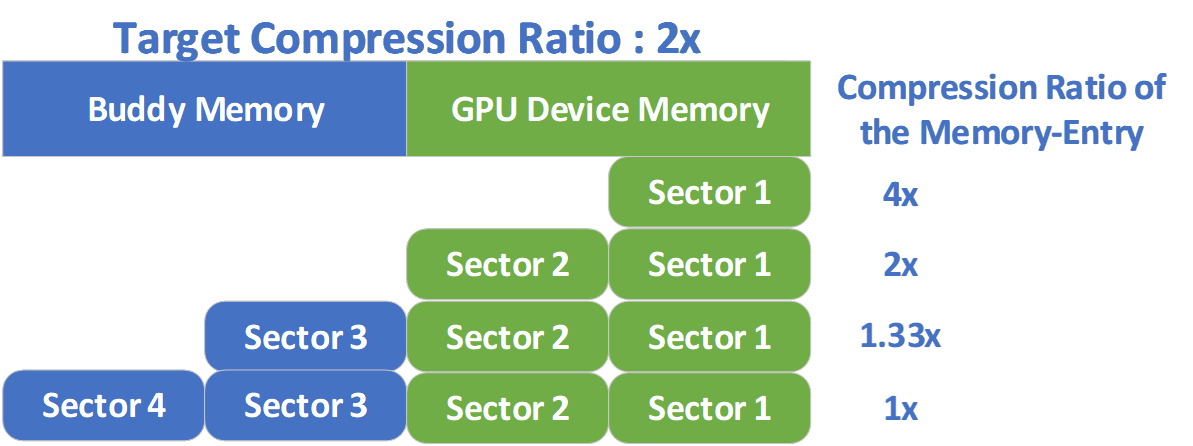}
  \caption{Depending on its data, a 128B memory-entry compresses to occupy from 1-4 sectors of 32B each. Here, the target compression ratio is 2x. If an entry does not compress to 2x, left over sectors are accessed from buddy-memory.}
  \label{fig:buddy_overview}
  \vspace{-10pt}
\end{figure}

\begin{figure*}[h]
\vspace{-1em}  
\centering
\subfloat[High-level overview of architectural changes/additions with Buddy Compression.] {
	\includegraphics[width=0.35\textwidth]{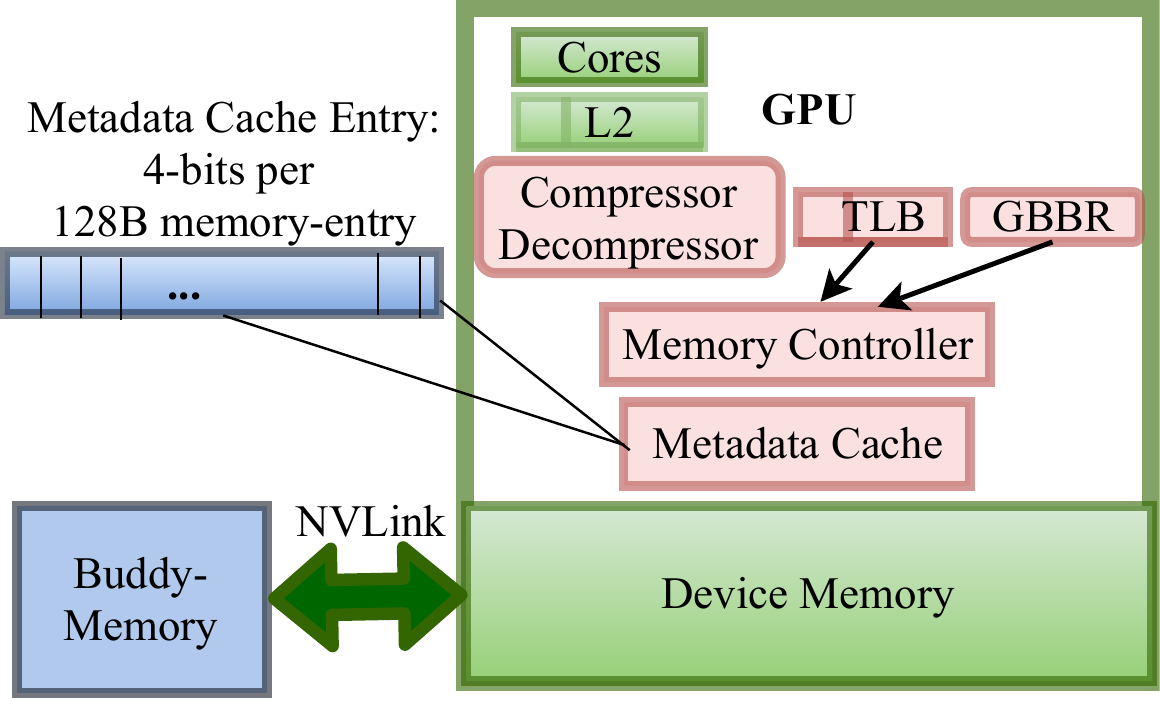}
	\label{fig:meta_overview}
}
\subfloat[Metadata cache hit rates with different sizes of total metadata cache] {
	\includegraphics[width=0.63\textwidth]{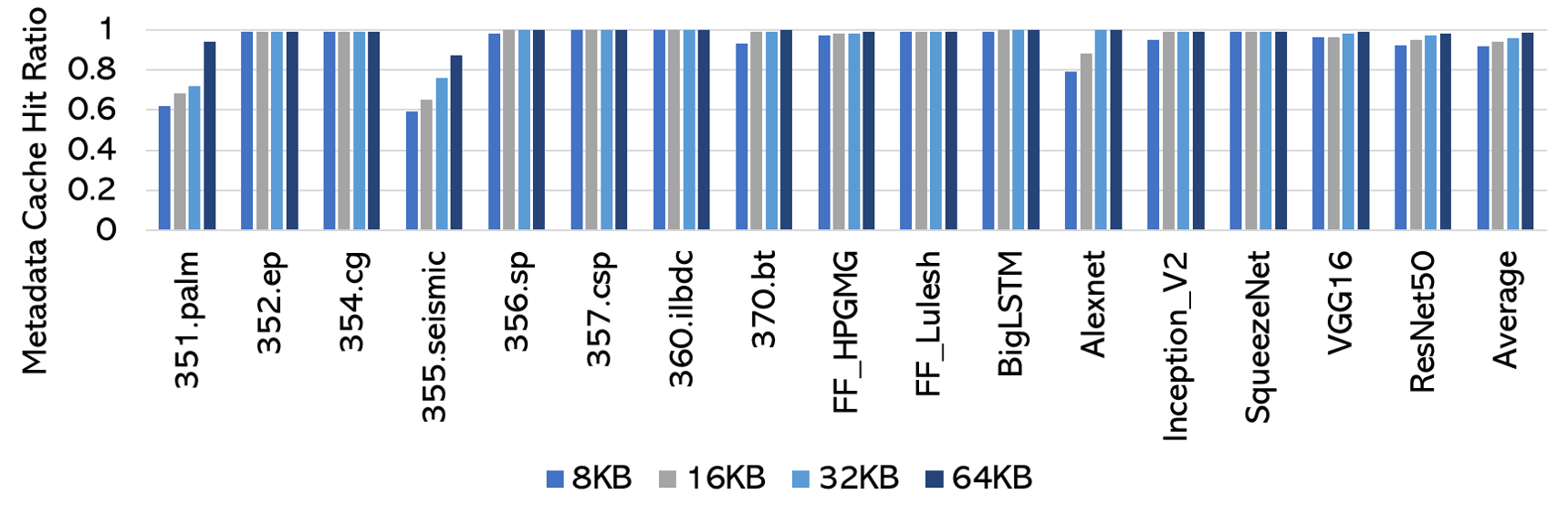}
	\label{fig:metadata}
}
\caption{Compression metadata handling and architecture}
\end{figure*}

Buddy Compression allows a programmer or DL framework to annotate the program
such that they take up less device memory than the allocation size.
For instance, if a user has 24GB of data and a GPU with only 12GB of memory
capacity, the data can be allocated with a target of 2x compression. This means that
only half of the full data size is allocated on the GPU device
memory. We use compression to
opportunistically fit data into this reduced device-resident allocation.
If a memory-entry does not compress sufficiently,
an NVLink-connected larger-but-slower memory is available as overflow storage.
The \textit{buddy-memory} is used as extended storage, with the
data being striped across at
128B memory-entry granularity. Data from compressible memory-entries is sourced
completely from GPU device memory, while incompressible memory-entries are
sourced from both device and buddy-memory.


As shown in Figure~\ref{fig:buddy_overview}, Buddy Compression stripes the data
using 32B sectors. This 32B sector size is
chosen to match the access granularity of GPU memory, specifically, 
GDDR5, GDDR5X, GDDR6, and HBM2-based GPUs.
For example, if an allocation targets a compression ratio of $2x$, the first two
sectors per 128B memory-entry are mapped to device memory, while the last
two are mapped to system memory. Therefore, if a memory-entry can
be compressed by $2x$ or more, it fits completely in device memory, and otherwise,
the rest of its data is saved in its fixed pre-allocated
spot in the buddy-memory.
The allowed compression ratios for this study are $1x$, $1.33x$,
$2x$ and $4x$ (4, 3, 2, or 1 sectors allocated in device memory).
These ratios are chosen to keep the sector interleaving simple
and avoid unaligned sector accesses. 

\mnew{
\textbf{Buddy-Memory Carve-Out Region.}
At boot time, the host system carves out a physically-contiguous chunk of its
memory for each GPU, dedicating this storage to be used as each GPU's buddy
memory. Those regions are then never accessed by the host, eliminating any
coherence issues. The buddy-memory size should correspond to the maximum target
compression ratio for the GPU. As an example, if the maximum target compression
ratio is $4x$, then the carve-out region should be $3x$ as large as GPU device
memory, in order to allow each memory-entry to have $3$ sectors in host memory (in the
worst case) and only $1$ on the GPU.

\textbf{Compression Metadata and Address Translation.}
Once the data is stored in compressed form, addressing it requires some additional translation and metadata.
This metadata informs us about 
\begin{inparaenum}[\itshape (i)]
\item the target compression ratio, 
\item whether or not a particular memory-entry was compressed to the target ratio, and
\item the address in buddy-memory to be accessed for memory-entries that did not compress to the target ratio.
\end{inparaenum}

A global base address for the buddy-memory carve-out region is stored in a Global Buddy Base-address Register (GBBR).
The page-table and TLBs are extended to store the information about whether the page is compressed or not, the target compression ratio, and the offset of buddy-page from the global base address.
This adds a total overhead of $24$ bits per page-table entry.
To know the actual compressed size of each 128B memory-entry, there are 4 bits of metadata per cache block, stored in a dedicated region of device memory, amounting to a $0.4\%$ overhead in storage. 
The metadata storage overheads of Buddy Compression are either comparable to, or less than those of previous works in memory compression in CPUs~\cite{compresso2018,lcp2013,mxt2001,rmc2005,Buri}.
Figure~\ref{fig:meta_overview} shows a very high-level view of the metadata setup and translation.
The simple GBBR-offset based addressing makes the overall translation mechanism very simple.

A \textbf{metadata cache} is used to avoid additional memory accesses each time memory is accessed.
Figure~\ref{fig:metadata} shows the metadata cache hit ratios as a function of the metadata cache size.
Most applications have high hit ratios.
We use a $4$-way $64KB$ metadata cache, that is split into 8 slices, 1 per DRAM channel.
Each metadata cache entry is $32B$, thereby causing a prefetch of metadata corresponding 63 neighboring $128B$ memory-entries on every metadata cache miss.
The metadata is assumed to be interleaved across the DRAM channels using the same hashing mechanism as regular physical-address interleaving.
}
\begin{figure*}
\vspace{-1em}
\scriptsize
\captionsetup[subfigure]{labelformat=empty,font=small,margin=0cm,labelfont={bf,up},textfont={up}}
  \centering
\begin{minipage}[b]{.08\textwidth}
    \centering
    \includegraphics[height=7.35cm, width=1.2cm]{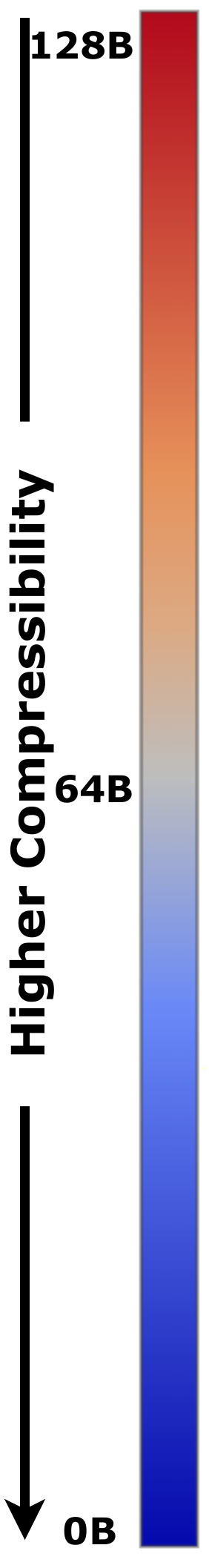}
\end{minipage}
\hspace{-5pt}
\begin{minipage}[b]{.9\textwidth}
    \centering
  \subfloat[351.palm] {
    \includegraphics[height=1.45cm, width=0.23\textwidth]{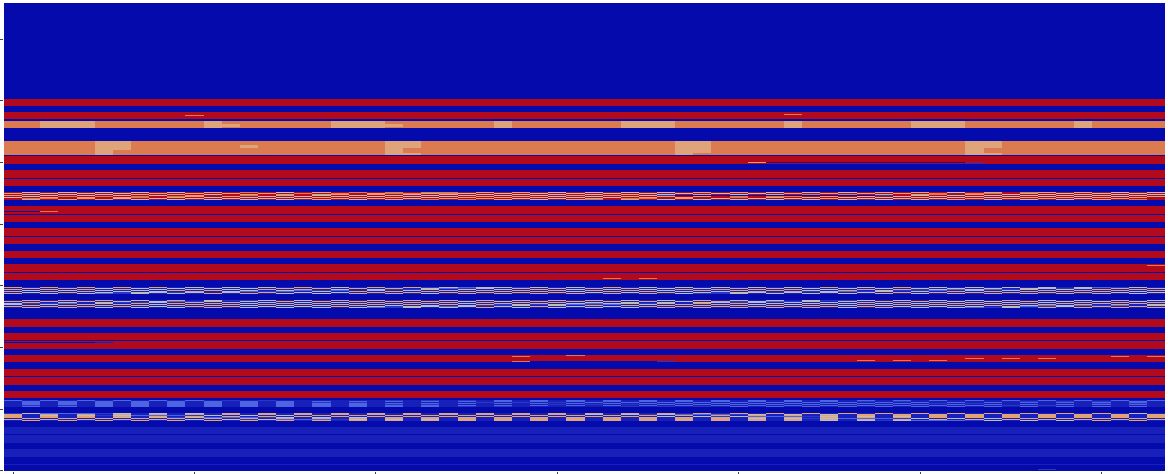}
  }
\hspace{-5pt}
  \subfloat[352.ep] {
    \includegraphics[height=1.45cm, width=0.23\textwidth]{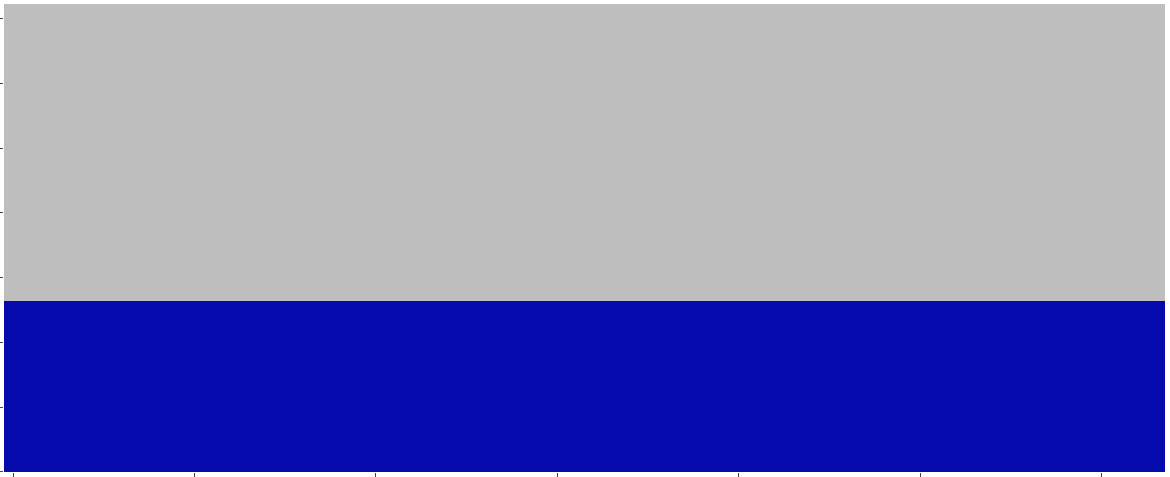}
  }
\hspace{-5pt}
  \subfloat[354.cg] {
    \includegraphics[height=1.45cm, width=0.23\textwidth]{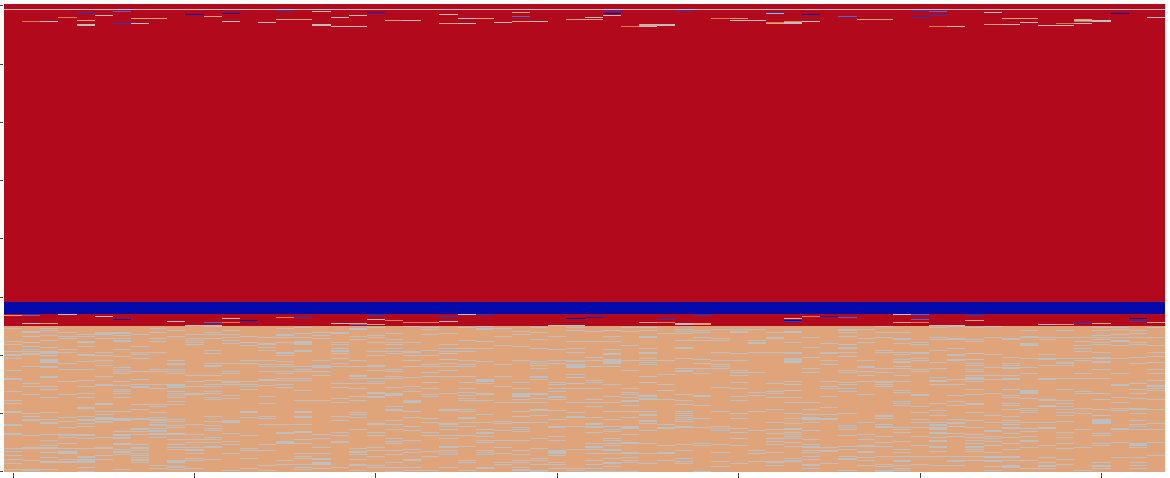}
  }
\hspace{-5pt}
  \subfloat[355.seismic] {
    \includegraphics[height=1.45cm, width=0.23\textwidth]{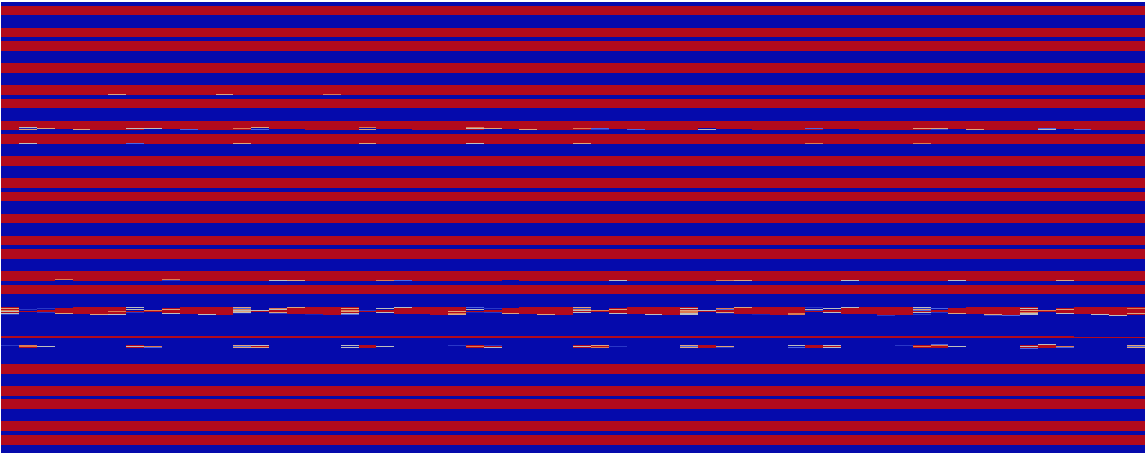}
  }
\vspace{-11pt}
  \subfloat[356.sp] {
    \includegraphics[height=1.45cm, width=0.23\textwidth]{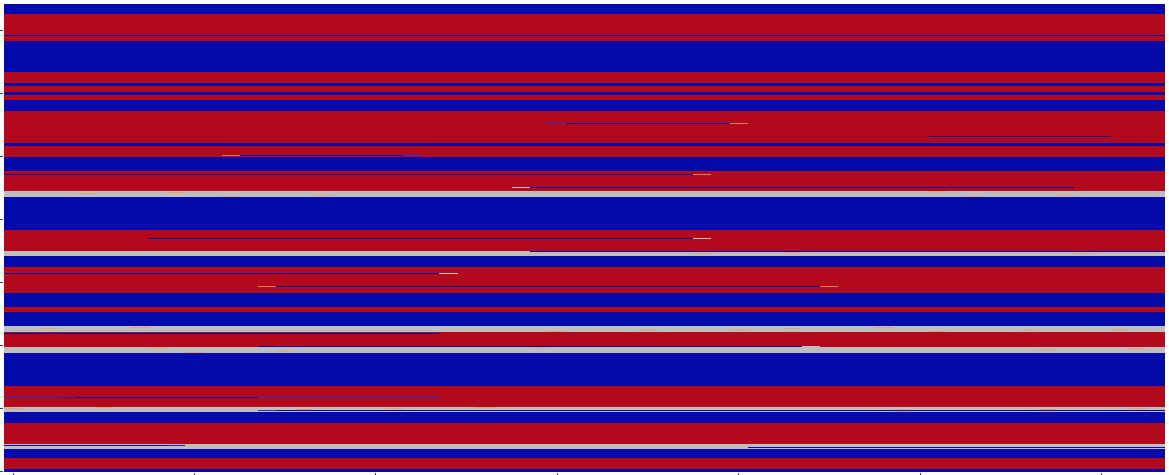}
  }
\hspace{-5pt}
  \subfloat[357.csp] {
    \includegraphics[height=1.45cm, width=0.23\textwidth]{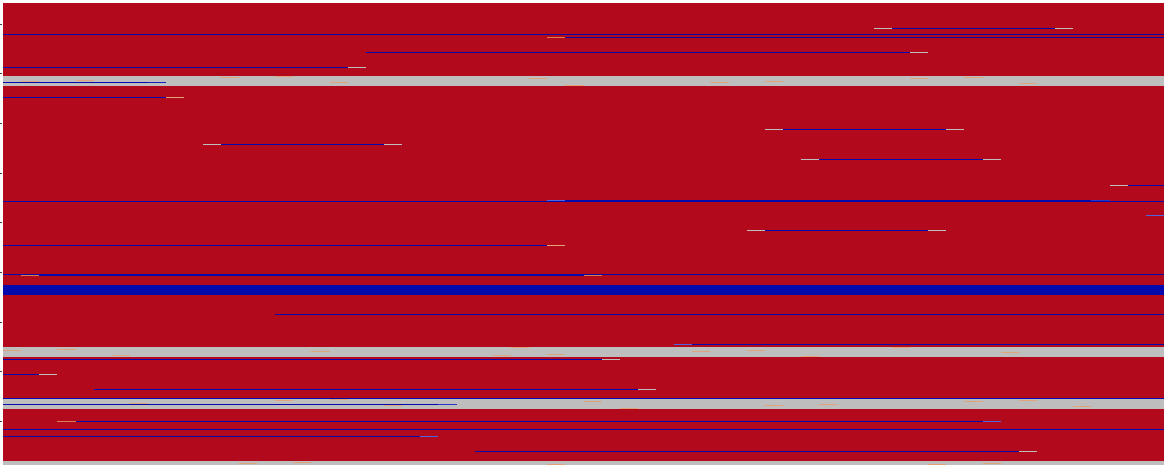}
  }
\hspace{-5pt}
  \subfloat[360.ilbdc] {
    \includegraphics[height=1.45cm, width=0.23\textwidth]{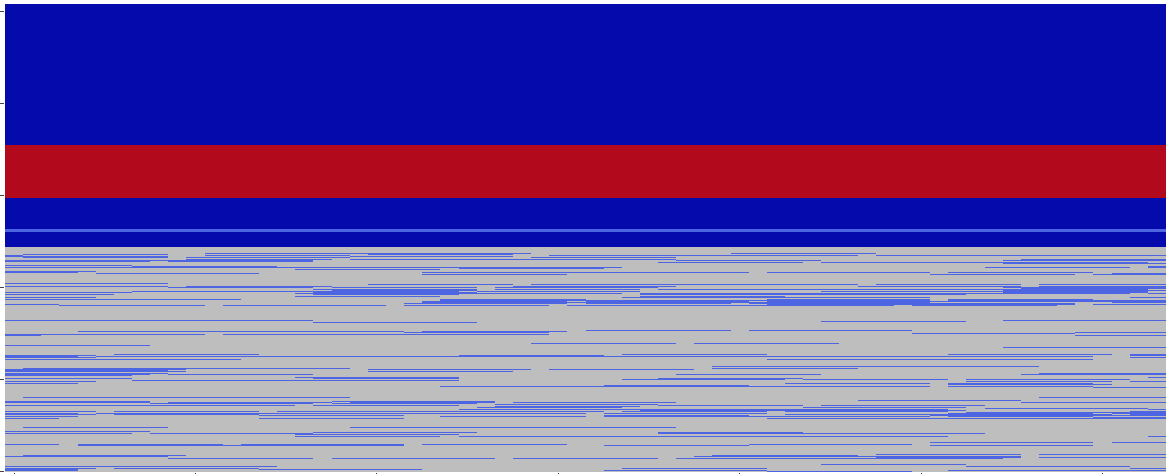}
  }
\hspace{-5pt}
  \subfloat[370.bt] {
    \includegraphics[height=1.45cm, width=0.23\textwidth]{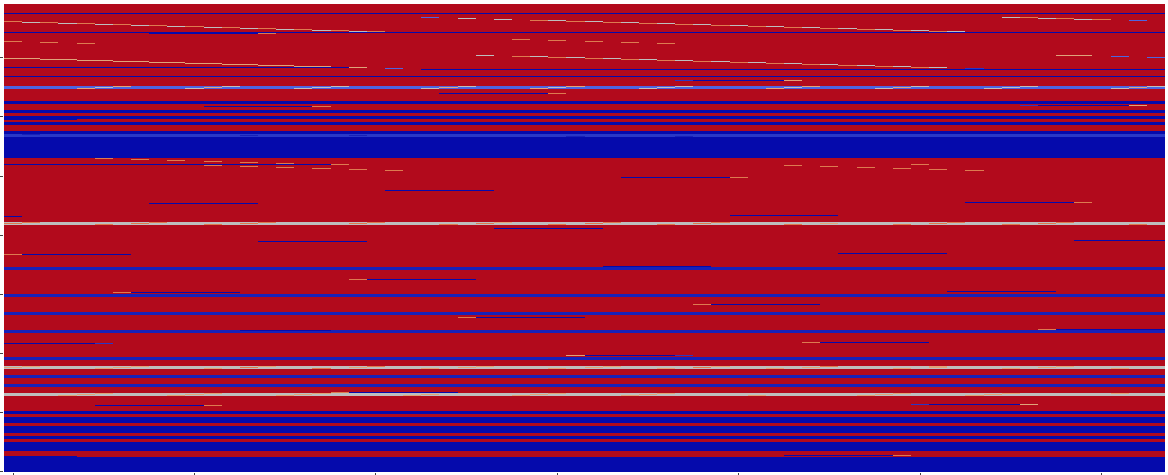}
  }
\vspace{-11pt}
  \subfloat[FF\_HPGMG] {
    \includegraphics[height=1.45cm, width=0.23\textwidth]{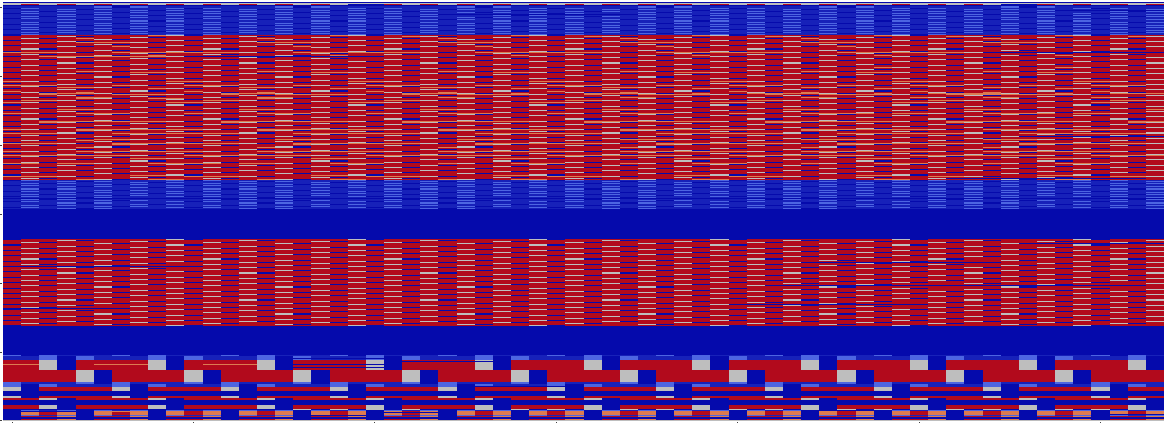}
  }
\hspace{-5pt}
  \subfloat[FF\_LULESH] {
    \includegraphics[height=1.45cm, width=0.23\textwidth]{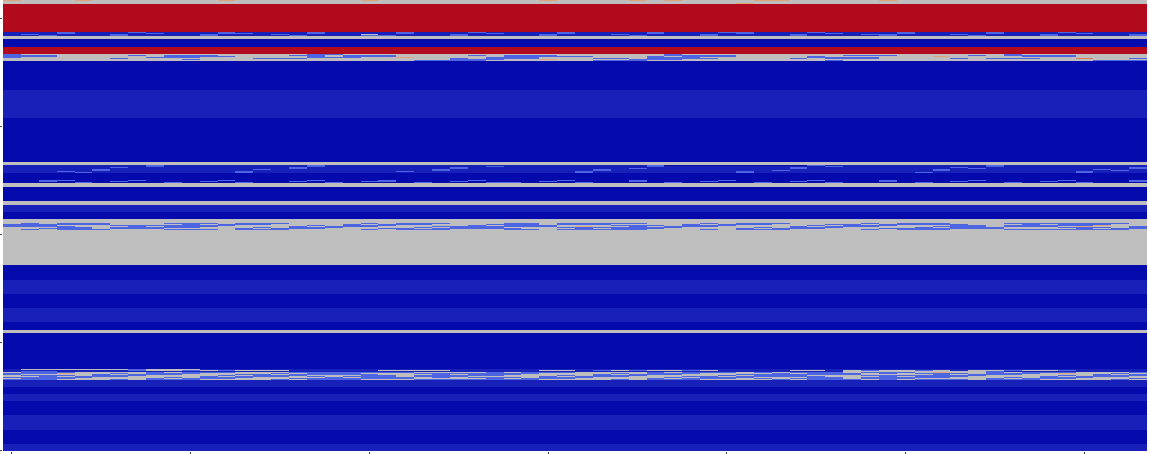}
  }
\hspace{-5pt}
  \subfloat[BigLSTM] {
    \includegraphics[height=1.45cm, width=0.23\textwidth]{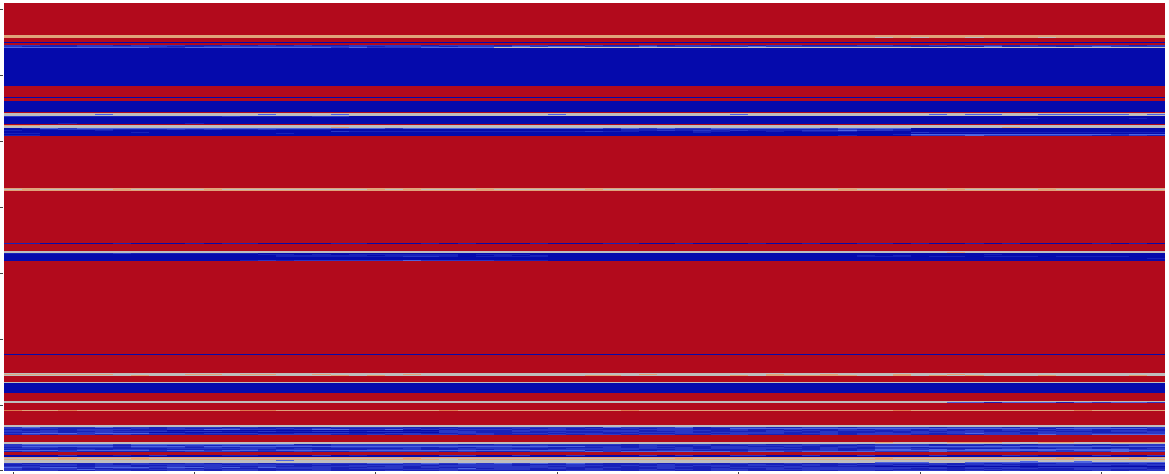}
  }
\hspace{-5pt}
  \subfloat[AlexNet] {
    \includegraphics[height=1.45cm, width=0.23\textwidth]{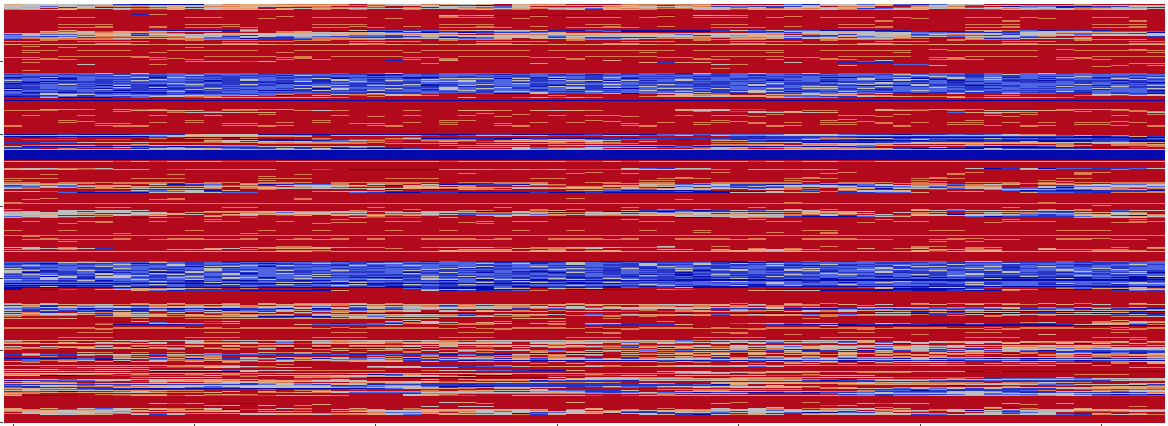}
  }
\vspace{-11pt}
  \subfloat[InceptionV2] {
    \includegraphics[height=1.45cm, width=0.23\textwidth]{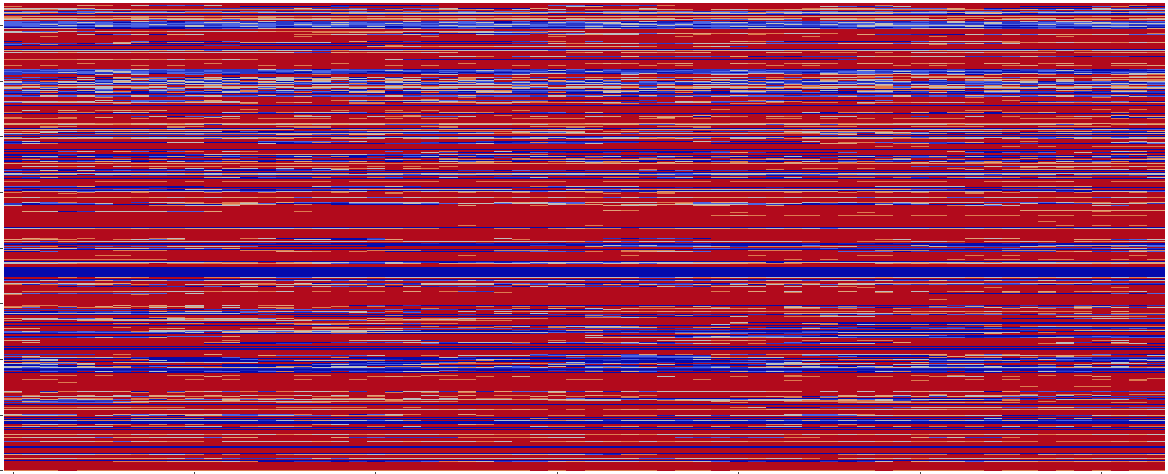}
  }
\hspace{-5pt}
  \subfloat[Squeezenetv1.1] {
    \includegraphics[height=1.45cm, width=0.23\textwidth]{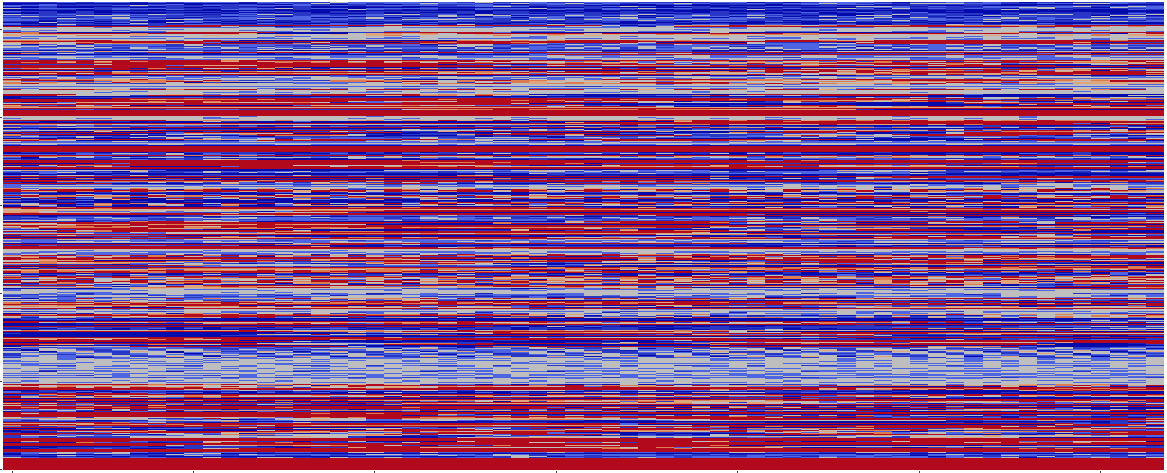}
  }
\hspace{-5pt}
  \subfloat[VGG16] {
    \includegraphics[height=1.45cm, width=0.23\textwidth]{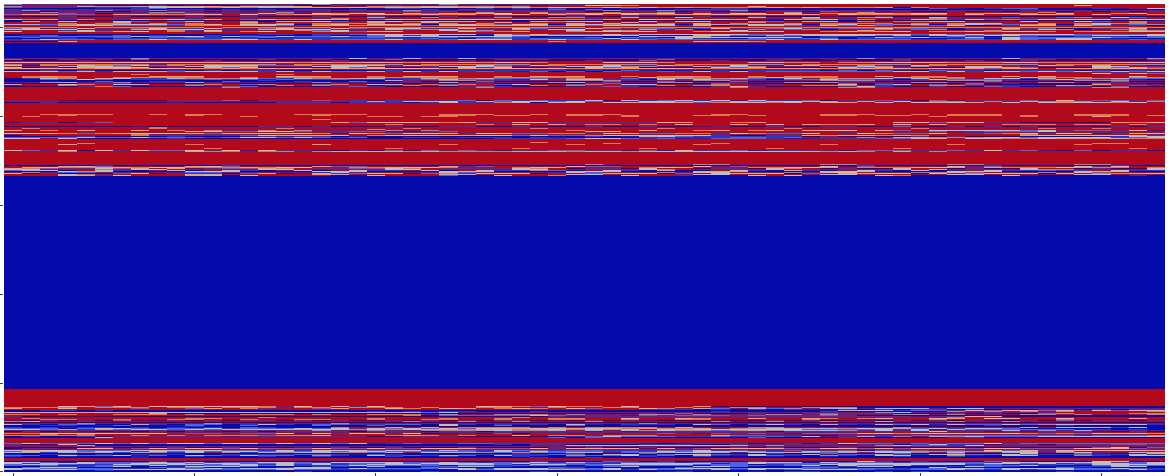}
  }
\hspace{-5pt}
  \subfloat[ResNet50] {
    \includegraphics[height=1.45cm, width=0.23\textwidth]{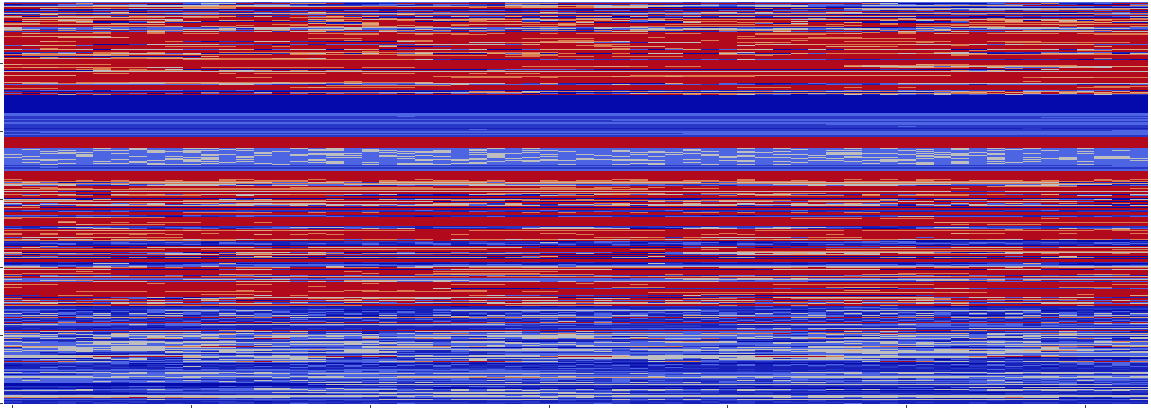}
  }
  \end{minipage}
\normalsize
  \caption{Spatial patterns of compressibility. Each plot is a heatmap of compressibility per 128B memory-entry for the allocated GPU memory. Each horizontal line is an 8KB page and pages are stacked vertically as per address.}
  \label{fig:spatial}
  \vspace{-1em}
\end{figure*}

\subsection{Benefits of the Buddy Compression Design}
{\bf No Page-Faulting Expense.} The immense parallelism of a GPU increases the
throughput of work done.  Driver-based page-fault handling, however, is remote
and non-distributed, making GPU page-faults during the runtime of a kernel
very expensive~\cite{tianhao_uvm}.
As data is written back to memory, its compressibility can decrease,
requiring new page allocations to store the same data. The page fault overhead
in GPUs makes reducing the compressed data movement a very important
directive. 
\textit{The uniqueness of the design lies in the fact that the
compressibility of each memory-entry affects only its own allocation, thereby
never having to cause page movement.}

{\bf Low Translation Overhead.} Memory bandwidth is a frequent bottleneck for GPUs. Accordingly, there
has been fruitful research on bandwidth compression of GPU main
memory~\cite{bpc2016,toggle2016}. Buddy Compression uses compression to amplify
\textit{both} the bandwidth and capacity of GPU memory. However, as discussed
earlier, compression-for-capacity requires additional metadata accesses
for translation into the compressed address space.
This makes reducing the metadata size and keeping translation simple important.
\mnew{Buddy Compression requires only 0.4\% metadata, and
since the carve-out region is contiguous in host memory, addressing into the buddy-memory is
offset-based, and trivial.

{\bf No Impact on Small Workloads.}
If the available GPU device memory is enough to allocate the memory required by the application,
the Buddy Compression can be disabled. In that case, the design does not affect
the performance at all.
}




\subsection{Reducing Buddy Compression Overheads} 

With the design of Buddy Compression, the obvious overhead comes from having to
access the slower buddy-memory in cases of unexpectedly low compression.

\textbf{Profiling for Target Compression Ratio.}
Choosing the right target compression ratio is important, since aggressive compression
ratios will lead to more memory-entries exceeding the allocated device
memory and requiring buddy-memory accesses. To choose the target compression
ratio, we use a simple profiling pass on a representative dataset.
For HPC workloads, the profiling pass is run using a smaller dataset, like the train
dataset for SpecAccel2.
For DL workloads, this profiling pass is run with a smaller batch size,
and can be embedded in the training platform, like PyTorch or TensorFlow.
Furthermore, the target ratios can be periodically updated for long running applications,
e.g., for DL training, the target ratio update can be combined with checkpointing in the
framework. In this paper, for simplicity, we consider a single static target compression
ratio throughout the run of the application.

\begin{figure*}
  \centering
  \includegraphics[width=0.85\textwidth]{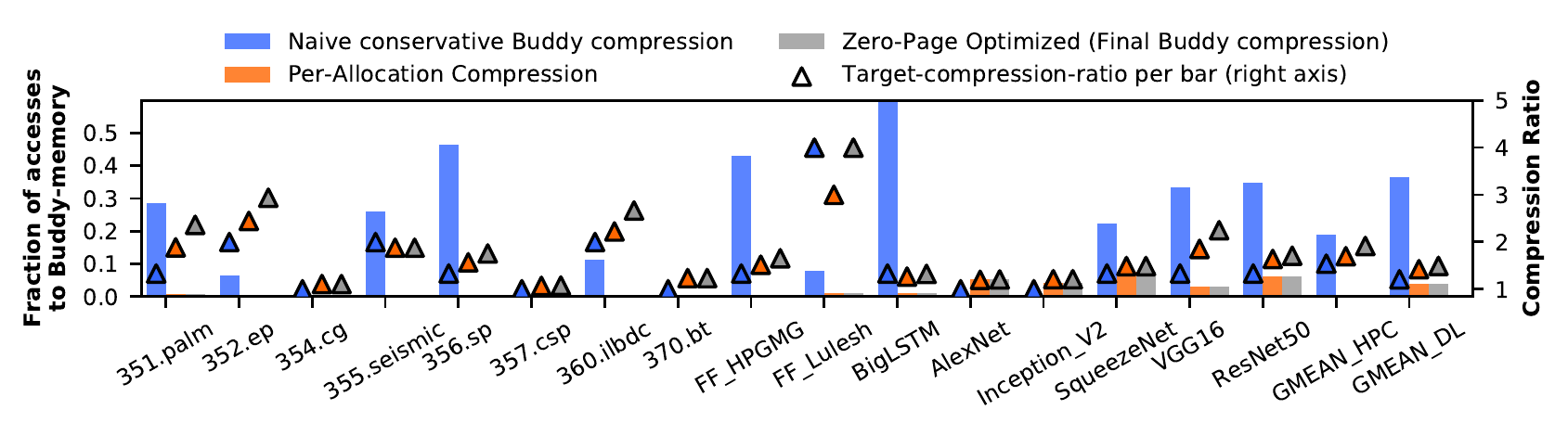}
  \vspace{-5pt}
  \caption{Sensitivity of the compression ratio and buddy-memory accesses to design optimizations.}
  \label{fig:Final_memAccesses}
  \vspace{-5pt}
\end{figure*}

\textbf{Annotation Granularity.}
The granularity with which the programmer annotates memory is also
important---the best annotation granularity depends on the spatial
characteristics of compressibility.
Naive Buddy Compression considers a single, conservative target
compression ratio for the whole-program.
As shown in Figure~\ref{fig:Final_memAccesses}, we find this granularity
to be too coarse.
The naive mechanism achieves an overall compression
ratio of $1.57x$ for HPC workloads, and $1.18x$ for DL workloads, with
$8\%$ accesses over the interconnect to the buddy-memory for HPC,
and $32\%$ for DL. The overall compression is low, and,
given that even the highest available bandwidth on the
interconnect (NVLink2, 150GBps) is 6x lower than the GPU device memory
bandwidth (900GBps), the overheads from these buddy-memory accesses are
high.
The desirable solution for us would be something that effectively lowers the
buddy-memory accesses, while maintaining high compression ratios. In order to
find such a solution, we present a deep dive into the detailed
compression data patterns in these workloads.


\begin{figure}
  \centering
  \includegraphics[width=0.8\columnwidth]{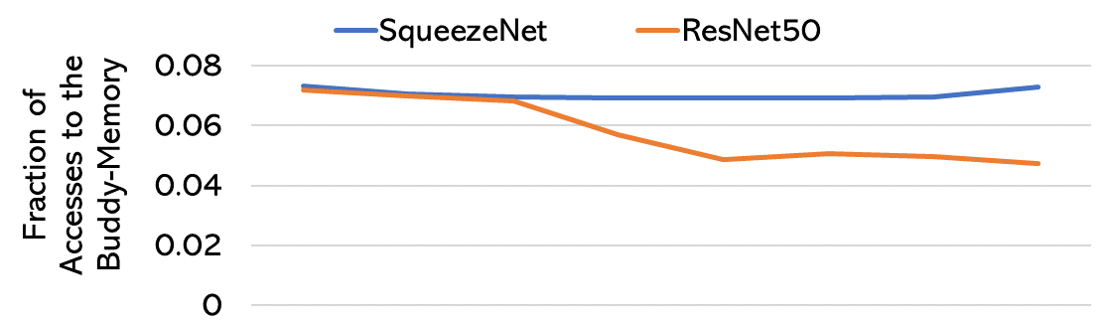}
  \caption{The fraction of buddy storage accesses over the execution of one iteration in DL training. We achieve a constant compression ratio of $1.49$ for $SqueezeNet$ and $1.64$ for $Resnet$.}
  \label{fig:ResNetSqueeze}
  \vspace{-10pt}
\end{figure}

\textbf{Understanding Compressibility Patterns.}
Figure~\ref{fig:spatial} shows a spatial plot (in the virtual address space) of
the compressibility of each workload's data. Each sub-plot is a spatial heat
map that shows the compressibility of each 128B memory-entry in the memory
allocated by each benchmark.  A colder color (blue), signifies high
compressibility and hotter color (red) shows low compressibility. The plot
is structured assuming $8KB$ pages, where each page has $64$ 128B
memory-entries laid along x-axis.  The y-axis is the total
number of pages in the memory of the benchmark. Figure~\ref{fig:spatial} shows
that the spatial locality of compressibility of data varied significantly varied across
benchmarks. While most HPC benchmarks have large homogeneous regions of similar
compressibility, the distribution is more random in DL workloads.  $FF\_HPGMG$ shows
specific patterns of compressibility that can directly be correlated with the
arrays of heterogeneous structs that are used in its allocation. Although the DL
workloads do not show the level of homogeneity that can be seen in HPC
workloads, there are still some mostly-red or
mostly-blue regions. Based on the insights from these
plots, we propose optimizations to the design of Buddy Compression.


\textbf{Per-Allocation Compression Targets.}
Figure~\ref{fig:spatial} shows that there are several regions that are mostly-red, or mostly-blue.
We find that the majority of these region boundaries overlap with $cudamalloc()$ boundaries.
A special allocation API for compressed regions allows us to capture this behavior and eliminate the futile effort of compressing the red regions.

During profiling, we periodically take snapshots of memory, to track the compression ratios per allocation.
At the end of profiling, we decide target compression ratios per allocation using heuristics to trade-off the compression ratio with the buddy-memory accesses.
The compression ratio chosen is conservative to minimize the buddy-memory accesses.
As an example, based on Figure~\ref{fig:overtime}, for $355.seismic$, for most allocations, the target ratio used will be $2x$, and not $7x$ or $6x$.
We use a static target compression ratio for the entire run of the application.
This is because a dynamic target compression ratio would require reallocating and moving around the pages, making the compression management more complicated and less performant, unless the applications are very long running and the overheads are amortized.

\begin{figure*}[tbh]
  \vspace{-1em}  
  \centering
  \includegraphics[width=0.85\textwidth]{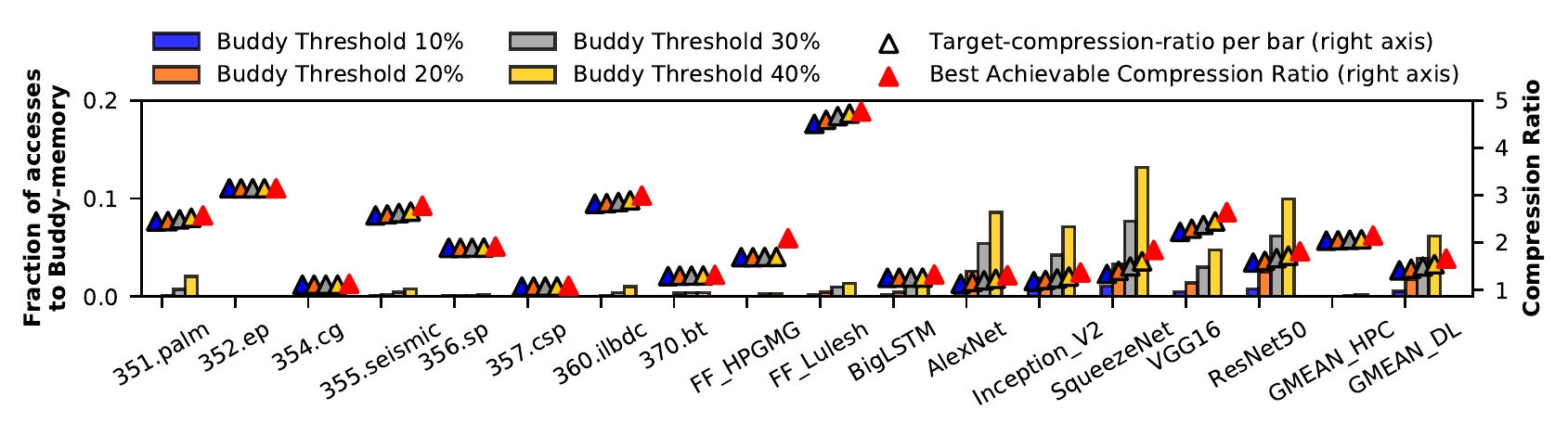}
  \vspace{-5pt}
  \caption{Sensitivity of the compression ratio and buddy-memory accesses to the Buddy Threshold parameter.}
  \label{fig:Buddy_Threshold}
  \vspace{-5pt}
\end{figure*}
\textbf{Buddy Threshold.}
Most benchmarks have regions that are highly homogeneous in their compressibility, making the per-allocation target ratio decision simple.
However, for benchmarks like $AlexNet$ and $ResNet50$, the regions are mixed in compressibility.
Therefore, target compression ratio decision in these cases, involve a trade-off between compression ratio and buddy-memory accesses.
We define a Buddy Threshold, that sets a limit on the fraction of memory-entries that require accessing the buddy-memory, per-allocation.
A higher Buddy Threshold achieves a higher compression ratio at the cost of more buddy-memory accesses, and hence, lower performance.
These buddy-memory accesses are calculated per target compression ratio, using a histogram of the static memory snapshots.

Figure~\ref{fig:Buddy_Threshold} shows the results from a sensitivity study of the Buddy Threshold ($10\%$ to $40\%$).
In addition to this, the figure shows the best achievable compression ratio assuming no constraints are placed on the buddy-memory accesses.
The bars in the figure show that the buddy-memory accesses remain very low for HPC benchmarks, due to their homogeneous regions.
For DL benchmarks however, the buddy-memory accesses are more frequent, and increase further as the buddy threshold is increased.
Similarly, the compression benefits from increasing the buddy threshold are mostly seen in DL benchmarks.
With the exception of $FF\_HPGMG$, we are able to achieve near-optimal compression, as can be seen in comparison with the black marker.
$FF\_HPGMG$, as discussed earlier has a peculiar striped compressibility pattern resulting from the struct it uses.
To capture the maximum compression, $FF\_HPGMG$ requires more than $80\%$ Buddy Threshold for most of its allocated memory region.
Another interesting scenario is seen in benchmarks $354.cg$ and $370.bt$.
Since these benchmarks mostly consist of incompressible data, without the per-allocation targets, Buddy Compression was unable to compress them at all.
However, with the per-allocation targets, we are able to compress them by $1.1x$ and $1.3x$ respectively.
Overall, since a $30\%$ Buddy Threshold achieves a good balance between the compression and buddy-memory accesses, we choose this for our final Buddy Compression design.

Since the target compression ratio remains constant, while actual data compressibility can change over time, these statically calculated buddy-memory accesses may not be similar across the run.
To investigate this, we observed the buddy-memory accesses across all the memory dumps, while maintaining constant target compression ratios.
Figure~\ref{fig:ResNetSqueeze} presents the results from $ResNet50$ and $SqueezeNet$, both of which have frequent changes in compression ratio per memory-entry, and have high accesses to buddy-memory to begin with.
We observe that the buddy-memory accesses do not change a lot over time.
This is because even though the individual memory-entries frequently change their compressibility, the changes are almost equal in both directions, making the overall effects small.
Furthermore, as mentioned earlier, for benchmarks that see large changes in their overall compressibility, like $355.seismic$, we avoid this challenge by choosing conservative target compression ratios.

\textbf{Special Case For Mostly-Zero Allocations.}\label{sec:zero_page}
Based on the spatial plots, we observe that there are areas in memory that
remain mostly-zero even across complete benchmark executions.  To capture the
capacity-expanding opportunity of such allocations, we add an aggressive
target compression ratio of $16x$ where we keep only $8B$ out of each $128B$ in
device memory.  Note that the only change involved here is an additional
encoding for page size in the TLB.


This optimization allows us to increase the compression ratio for benchmarks with large highly-compressible regions, for example, $352.ep$, and $VGG16$.
Note that this optimization does not have much impact on the buddy-memory accesses, since such compressible data would always fit in device memory.
Figure~\ref{fig:Final_memAccesses} shows the impact of this optimization.
For HPC benchmarks, the compression ratio goes up from $1.7x$ to $1.9x$, while for DL, from $1.42x$ to $1.5x$.

For this optimization, it is important to identify allocations that are mostly zero, and remain so for the entirety of the run of the benchmark, unless there is a periodic update of target compression ratio.
The profiler marks the regions that can be compressed with this optimization, such that the overall compression ratio is still under 4x, limited by the buddy-memory carve-out region.

\textbf{Possible Optimization for Metadata Access.}
We note that on a metadata cache miss, both the device-memory data and its metadata can be accessed in parallel, since the metadata only informs us about the buddy-memory part of data.
We do not, however access the buddy-memory in parallel, since the buddy-memory accesses are rare on average (Figure~\ref{fig:Final_memAccesses}).

\subsection{Final Design}
Buddy Compression uses a Buddy Threshold default of 30\%, a metadata cache of 4KB per DRAM channel, and a buddy-memory region of size 3x of the GPU device memory, to support a 4x maximum compression ratio.
The application is first profiled with a smaller dataset, during which our tool periodically calculates a histogram of compressed memory-entries per allocation.
At the end of profiling, the tool reports the target compression ratios, which are then used by the DL platform, or the HPC user to annotate cudamalloc API calls, enabling running a larger dataset without the overheads of Unified Memory.
Figure~\ref{fig:Final_memAccesses} shows the compression ratio and buddy-memory accesses for the final design.
We achieve $1.9x$ compression for HPC and $1.5x$ compression for DL workloads.
The average buddy-memory accesses are $0.08\%$ for HPC and $4\%$ for DL workloads.

\section{Performance Evaluation}
We have already presented results regarding buddy-memory accesses and compression ratios from Buddy Compression in Figure~\ref{fig:Final_memAccesses}.
In this section, we first discuss the performance impact of Buddy Compression relative to an ideal large-capacity GPU, followed by a comparison to UM-based oversubscription.
We then present a case-study of DL training to estimate the performance benefits from increased capacity.
\subsection{Methodology}
\label{sec:meth}

\textbf{Workloads.}
As previously described, we evaluate Buddy Compression's effectiveness on 
workloads from the SpecAccel~\cite{specaccel} and FastForward benchmark suites 
for HPC workloads. We collect a representative trace from each benchmark while 
running the reference datasets. Each trace contains $1-9$ billion warp 
instructions and corresponds to the dominant kernel of each benchmark at a 
point in execution that exhibits the average compression ratio for that entire 
benchmark execution~\cite{CompressPoints}. For DL, we use a set of 5 convolutional neural networks: 
AlexNet~\cite{alexnet}, Inception v2~\cite{inception}, 
SqueezeNetv1.1~\cite{squeezenet}, VGG16~\cite{vgg16}, and 
ResNet50~\cite{resnet50}, all of which were run under the Caffe~\cite{caffe} 
framework with the ImageNet~\cite{imagenet_cvpr09} dataset. Additionally we 
consider a long short-term memory network, BigLSTM~\cite{Jzefowicz2016}, which 
is a 2-layer LSTM with a 8192+1024 dimensional recurrent state in each of the 
layers and uses the English language model. The traces for the
DL training workloads span one full training iteration.
\begin{figure}[b]
\vspace{-1em}
\centering
\hspace{-1em}\includegraphics[width=0.9\columnwidth]{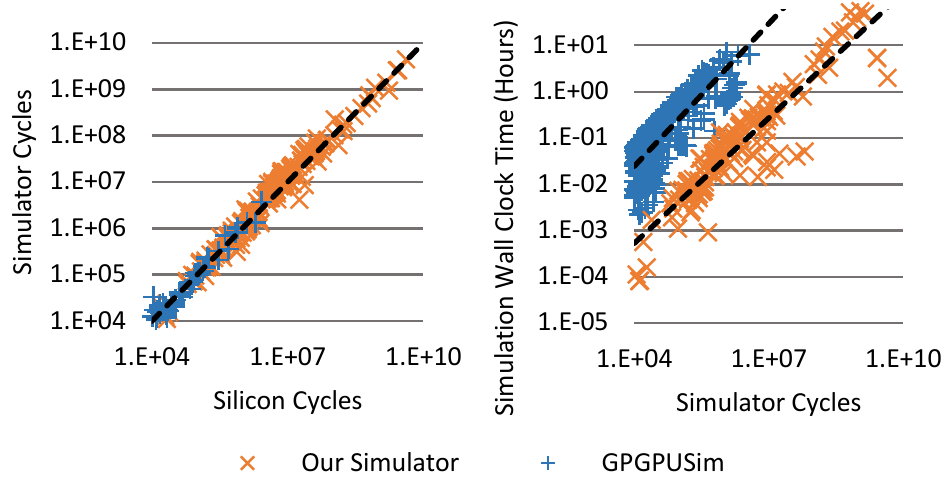}
\caption{Our simulator correlates with a V100 GPU (left, with slope=1 line). It is two orders of magnitude faster than GPGPUSim~\cite{gpgpusim}, enabling longer programs (right, linear regression lines).}
\label{fig:NVAS_corr}
\vspace{-5pt}
\end{figure}
%

\begin{table}
  \caption{Performance simulation parameters.}
  \label{tab:eval_config}
\begin{center}
\footnotesize
    \begin{tabular*}{\columnwidth}{@{\hspace{4pt}}M{0.065\textwidth}@{\hspace{0.5em}}M{0.43\textwidth}@{}}
\toprule
\multirow{2}{*}{Core}
            & 1.3 GHz; 2 greedy-then-oldest warp schedulers per SM \\
			& Max 64 32-thread warps per SM \\
\addlinespace[0.33em]
\multirow{3}{*}{Caches}
            & 24KB private L1/texture cache per SM, 128B lines \\
			& 64KB dedicated scratchpad per SM, \\
            & 4MB shared L2, 32 slices, 128B lines, 16 ways \\
\addlinespace[0.33em]
\multirow{2}{*}{Off-Chip}
            & 32 HBM2 channels at 875MHz (900 GBps)                     \\
            & 6 NVLink2 bricks (150 GBps full-duplex\textsuperscript{*})                     \\
\addlinespace[0.33em]
\multirow{2}{*}{Buddy}
            & 4KB\textsuperscript{*} metadata cache per L2 slice, 128B lines, 4 ways  \\
            & Compression/Decompression latency = +11 cycles \\
\bottomrule
\addlinespace[0.1em]
\mbox{\hspace{0.5em}* These parameters are swept in later parts of the evaluation.} &
\end{tabular*}
\end{center}
\vspace{-15pt}
\end{table}
\normalsize

\begin{figure*}[t!]
\vspace{-1em}  
\centering
\includegraphics[width=0.85\textwidth]{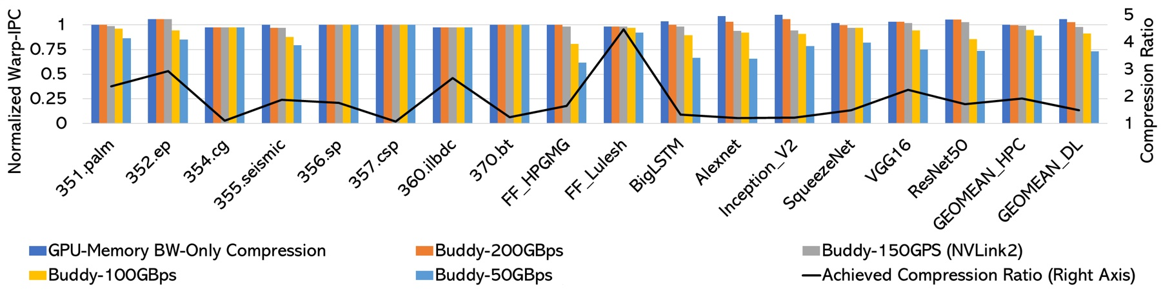}
\caption{The performance impact of compression, not accounting for capacity benefits. Systems with different link bandwidths are evaluated (showing unidirectional full-duplex bandwidths), with results normalized to a system with unlimited memory and a 150GBps interconnect.}
\label{fig:perf}
\vspace{-5pt}
\end{figure*}

\textbf{Simulation Infrastructure.} We use a dependency-driven GPU
performance simulator, similar to the one used by Arunkumar et
al. and others~\cite{mcm_gpu,milic2017beyond,young2018combining}.
We configure the simulator based on publicly information about
NVIDIA's P100 Pascal GPU~\cite{nvidia_pascal} and the interconnect
characteristics of recent Volta GPUs~\cite{nvidia_volta}
(\tab{tab:eval_config}). Non-public microarchitectural details are configured using
microbenchmark results from Jia et al.~\cite{jia_dissecting_2018}.  Each SM is modeled as an
in-order processor with greedy-then-oldest warp scheduling. We model a
multi-level cache hierarchy with private L1 caches and a shared sectored L2
cache with 128B lines and 32B sectors.  Caches are banked to provide the
necessary parallelism to saturate DRAM bandwidth. We model software-based cache
coherence in the private caches, similar to state-of-the-art GPUs. The memory
system consists of 32 HBM2 channels and the GPU is connected to the system with
6 NVLink2 bricks. 

We conservatively model decompression latency as 11 DRAM cycles, as discussed
in prior work~\cite{bpc2016}. Unless explicitly noted, the default metadata
cache configuration is 4-way set associative 4KB per L2 slice. Additionally, to
separate out its performance impact, we also evaluate bandwidth-only
interconnect compression between the L2 cache and device memory. Such
compression does not increase the effective memory capacity, but it can increase the
bandwidth between L2 cache and memory without requiring any metadata or
buddy-memory accesses. 

Figure~\ref{fig:NVAS_corr} (left) shows that our simulator correlates well
(correlation coefficient 0.989) against the total cycles spent on a real V100
GPU across a wide variety of benchmarks across a number of domains.\footnote{We
show simulator correlation results with a slightly different configuration than
is used for evaluation, in order to also be able to show comparable GPGPUSim
results. The P100 configuration used in the paper also correlates well with
silicon.} Corresponding numbers from GPGPUSim (correlation coefficient 0.948),
a widely-used academic simulator, are also shown. Our motivation in using a
proprietary simulator comes from the two orders-of-magnitude speed benefit
shown in Figure~\ref{fig:NVAS_corr} (right), which enables us to simulate
larger and more realistic workloads.

\subsection{Performance Relative to an Ideal GPU}
Apart from increasing the memory capacity, Buddy Compression can affect the performance of the system in the following ways:
\begin{inparaenum}[(i)]
\item Buddy-memory accesses can cause a performance hit.
\item Decompression latency can hinder performance.
\item Metadata cache misses can cause additional requests to device memory.
\item It has two conflicting effects on the effective bandwidth from L2 cache to device memory.
First, since compression is done at the cache-block granularity, the minimum L2 fill granularity is no longer a single 32B sector.
Instead, the complete cache-block is transferred to L2 upon a load access to a compressed memory-entry. This may result in over-fetch for fine-grained accesses, squandering device memory bandwidth.
However, for compressible workloads with high locality, compression allows an increase in effective bandwidth because cache-blocks can be fetched with fewer memory accesses.
\end{inparaenum}

We evaluate Buddy Compression alongside bandwidth-only compression that
compresses the data being transferred between L2 cache and device memory.
We also sweep the buddy-memory interconnect bandwidth from 50 to 200GBps on full-duplex connection,
where 150GBps represents NVLink2.
The results are shown in Figure~\ref{fig:perf}.
Bandwidth-only compression achieves an overall speedup of $5.5\%$.  Most of
this speedup comes from the DL training workloads.  This is because of
the regular, streaming memory accesses of these workloads, which are
essentially performing matrix multiplications. Since most of their memory
accesses are coalesced to access all sectors in each cache-block, bandwidth
compression achieves higher effective bandwidth by requiring fewer packets per
request. On the other hand, the HPC applications $354.cg$ and $360.ilbdc$
experience slowdowns with bandwidth compression.  This is because of the random
and irregular memory access pattern of these benchmarks.  Most of their memory
accesses require only one sector. However, bandwidth compression leads to a
full cache-block transfer of any compressible data, potentially lowering the effective
bandwidth for random accesses.  $FF\_Lulesh$ experiences a slowdown despite having a regular memory
access pattern. We find the reason behind this to be the
compression and decompression latency, which both lie on the critical path for
bandwidth compression.

Buddy Compression introduces additional overheads on top of bandwidth
compression, in the form of metadata cache misses and buddy-memory accesses.
Figure~\ref{fig:perf} shows that while an interconnect bandwidth of 200GBps
still achieves a 2\% average speedup using Buddy Compression, all lower
interconnect bandwidths experience some slowdown relative to the ideal
large-capacity GPU that serves as the baseline.  Note that the performance
benefits from a larger memory capacity are not accounted for in these
experiments.

The benchmarks $351.palm$ and $355.seismic$ experience slowdown due to a higher
metadata cache miss rate, as can be seen from Figure~\ref{fig:metadata}. Since
the other benchmarks have high metadata cache hit rates, metadata accesses do
not have a discernible impact on their performance.

Most HPC benchmarks have rare buddy-memory accesses
(Figure~\ref{fig:Final_memAccesses}), leading to negligible slowdowns with a
high bandwidth interconnect. However, when the interconnect bandwidth is
reduced, even these 1\% accesses from buddy-memory can cause a considerable
slowdown of bandwidth-sensitive applications like 352.ep and 355.seismic. Note
that FF\_HPGMG has host-memory accesses in its native form, due to synchronous
copies from host to device. Therefore, lowering the link bandwidth shows a
drastic impact on its performance (since all results are normalized to a
baseline system without compression and a 150 GBps interconnect).
 

DL training workloads have a higher percentage of buddy-memory accesses, as can
be seen in Figure~\ref{fig:Final_memAccesses}. These buddy-memory accesses are
caused by a lack of compression locality in the workloads. For example,
AlexNet requires accesses to $5.4\%$ of memory locations to go to buddy-memory,
leading to a $6.5\%$ slowdown relative to ideal (with a 150GBps full-duplex
interconnect). This is because of the difference in the bandwidth available
from device memory vs.~buddy-memory, which in our setup is $900GBps$ and
$150GBps$. Performance degenerates quickly as this disparity grows, with the 50GBps
full-duplex connection seeing a 35\% slowdown.


These results show that recently-developed high-speed GPU interconnects are an
enabling technology for Buddy Compression. The slowest link we evaluate (50
GBps full-duplex) is still faster than the most recent PCIe generation (x16
PCIe4.0, providing 32GBps full-duplex bandwidth) yet it suffers from more than
20\% average slowdown relative to the ideal GPU. However, using high bandwidth
interconnects such as NVLink2 (150Gbps full-duplex) enables Buddy Compression
to come within $1\%$ of the performance of the large-capacity GPU on HPC
benchmarks, and within $2.2\%$ of ideal on DL training workloads.

\subsection{Comparison with Unified Memory}

\begin{figure}[tbh]
  \centering
  \includegraphics[width=0.95\columnwidth]{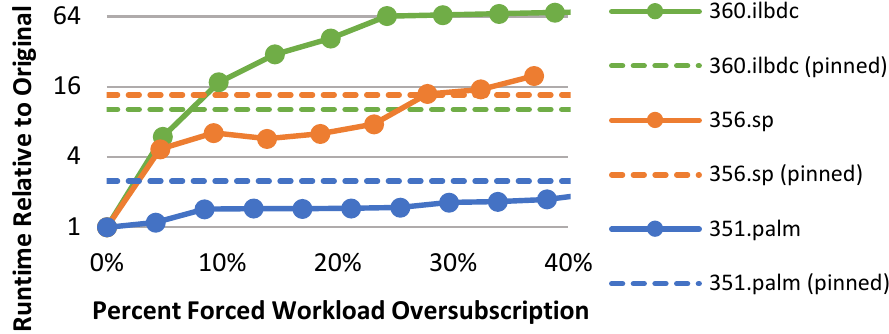}
    \caption{Measured overheads of using UM oversubscription. A Power9 CPU is connected via 3 NVLink2 bricks (75 Gbps full-duplex) to an NVIDIA V100 GPU. Dotted lines show the performance when all allocations are in the host memory.}
  \label{fig:UVM}
  \vspace{-1.5em}
\end{figure}

Faithfully comparing the performance of Buddy compression to Unified Memory in
simulation is not feasible due to the complex host-driver interactions and page
migration policies implemented within UM. Instead we choose to understand UM
performance in oversubscription scenarios on real hardware.
Figure~\ref{fig:UVM} shows measured performance of three applications on an IBM
Power9 system, connected to a Tesla V100 GPU via NVLink2 (3 bricks, 75 GBps
full-duplex bandwidth). We use SpecAccel applications with the \textit{managed}
PGI compiler flag, and force varying levels of oversubscription through an
interposer that hogs GPU memory at application startup. We also run the
applications using a compiler flag to pin all allocations in host memory,
showing the slowdown in dotted lines. Our results indicate that UM migration
heuristics often perform worse than running applications completely pinned in
host memory; perhaps because UM was primarily intended for the ease of
programming and has not yet been tuned for high-performance memory
oversubscription. Previous work~\cite{uvm_hpgmg,tianhao_uvm} supports our
observation that the slowdown due to UM oversubscription can be excessive
without more extensive hardware support. Figure~\ref{fig:perf} shows that Buddy Compression
suffers from at most 1.67x slowdown for these programs when oversubscribing by
50\%, even with a conservative 50 GBps NVLink speed. This indicates that it is
a better alternative to high performance memory over subscription than
software-based UM.


%

\subsection{Case Study: DL Training Benefits from Increased Memory Capacity}\label{sec:dl_stuff}
\begin{figure}[tp]
\vspace{-1em}  
\centering
\subfloat[Memory footprint of DL workloads as a function of batch size. (PyTorch, Titan Xp)] {
	\includegraphics[width=0.39\textwidth]{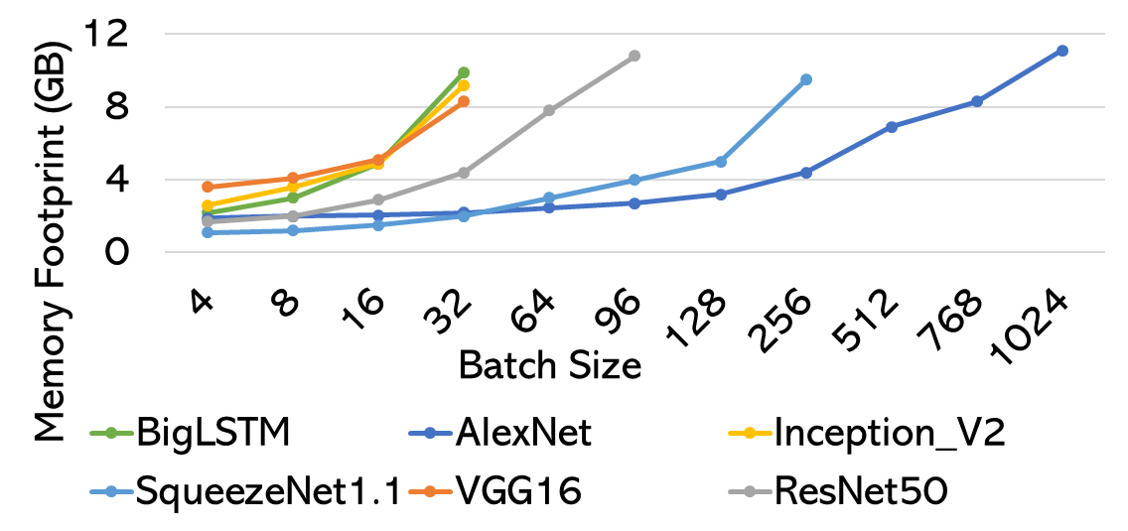}
	\label{fig:DL_footprint}
}
\vfill
\subfloat[Projected speedup in images per second as a function of mini-batch size.] {
	\includegraphics[width=0.39\textwidth]{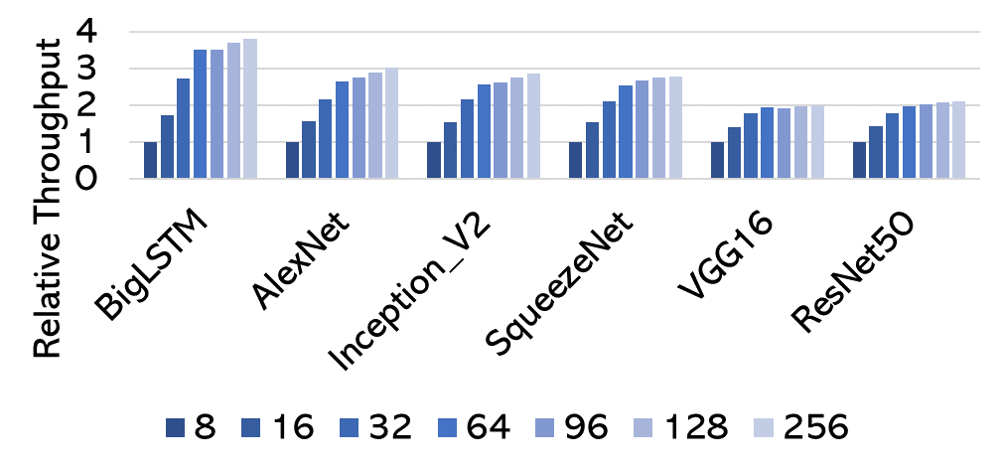}
	\label{fig:DL_sweep}
} 
\vfill
\subfloat[Projected speedup in images per second by using Buddy Compression to achieve a larger batch size.] {
	\includegraphics[width=0.39\textwidth]{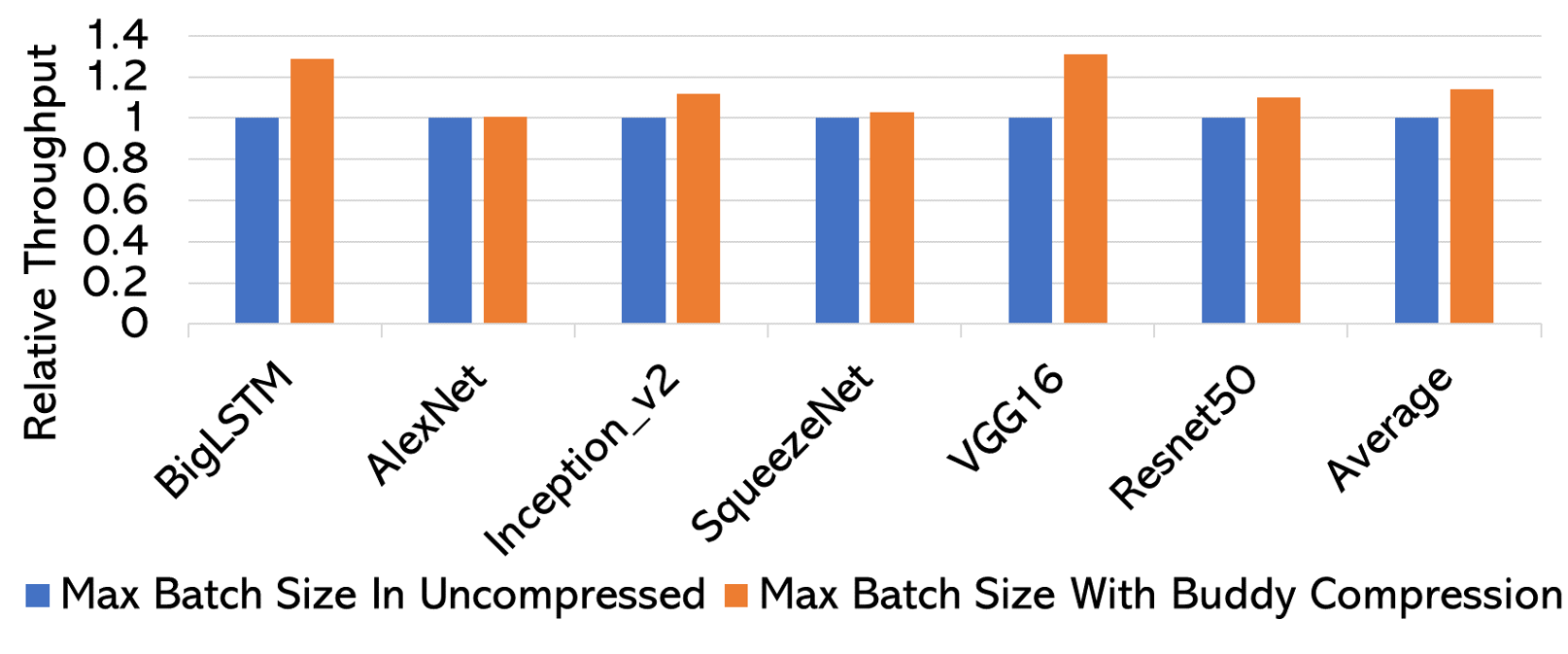}
	\label{fig:DL_fps}
}
\vfill
\subfloat[Validation accuracy with different mini-batch sizes. ResNet50 is trained until 100 epochs with CIFAR100.] {
	\includegraphics[width=0.39\textwidth]{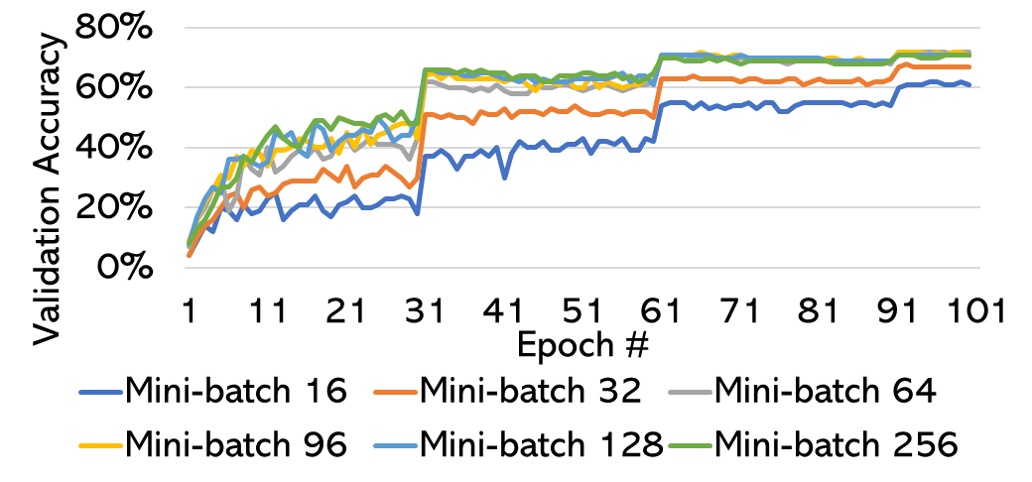}
	\label{fig:DL_train}
}
\vspace{5pt}
\caption{Impact of increasing mini-batch size on DL training}
\vspace{-1em}
\end{figure}

Thus far, we have compared the performance of Buddy Compression to an uncompressed, large-memory baseline (Figure~\ref{fig:perf}).
This excludes the benefits of having access to a larger-memory, thus ignoring the main purpose of Buddy Compression.
In the case of HPC benchmarks, a larger memory enables solving a larger problem. Such benefits are important yet difficult to quantify.
Accordingly, we instead perform a case-study on DL training workloads to quantify the performance benefits from compression.

Stochastic Gradient Descent (SGD) is widely used to update weight values during DL network training.
SGD iterates repeatedly through the training dataset, optimizing and updating the model each iteration. Updates depend on hyperparameters (such as the chosen learning rate) and dynamically react as the classification accuracy increases.
The entire dataset is divided into mini-batches, and each iteration goes through one mini-batch and updates the model.
\mnew{While there is an ongoing debate concerning the utility of large mini-batches across all domains, they can help regularize and improve convergence in many cases, as evidenced by~\cite{megdet2017,BERT}.}


\textbf{Memory Footprints of DL Workloads.} The memory footprint of a network
during training depends on the mini-batch size. Larger mini-batch sizes require
a larger part of the dataset to reside in device memory, along with more
intermediate data (activations and gradients). Figure~\ref{fig:DL_footprint}
shows the memory footprint of each of our DL training workloads as the
mini-batch size is increased. The sizes are increased up to the maximum size
that a Titan Xp GPU can support (12GB device memory). Initially there is not
much difference as the batch size is doubled.  Eventually, however, the memory
footprint grows almost linearly with increasing mini-batch size.  This
transition point depends on the size of the network parameters, which do not
vary with mini-batch size. For example, for AlexNet, the network parameters
are a large portion of the overall memory consumption due to the three large
fully-connected layers and relatively few (five) convolutional layers.
This leads to a later transition point for AlexNet at a batch-size of 96; all
other tested networks transition to an increasing memory footprint at a batch
size of 32 or below.

\textbf{Performance Impact of Larger Mini-Batches.}\label{sec:dl_proj}
A larger batch size is beneficial for DL
training~\cite{Jzefowicz2016,megdet2017}, because it allows more work to be
done per iteration, leading to higher resource utilization.
We use an analytical model very similar to~\cite{paleo2017,delta2019} to project this,
since we cannot collect traces for DL execution with memory capacity requirements
that are larger than current GPU's capacity.
We extensively validate the model and its projections highly correlate with a range of existing commercial GPUs.
Figure~\ref{fig:DL_sweep} shows the projected speedup for each network as the
mini-batch size is increased. It is generated using a detailed analytical model
of deep learning training efficiency (Section~\ref{sec:meth}).  As shown in the
figure, increasing the mini-batch size leads to higher speed in terms of frames per
second.  This effect, however, is only seen until the mini-batch size is large
enough to utilize most of the GPU resources. After the point of full GPU
utilization, the effect plateaus.


Buddy Compression allows us to fit a larger mini-batch into GPU memory.
Figure~\ref{fig:DL_fps} shows the relative speedup projected by our model for
this larger mini-batch size over a baseline GPU with 12GB of device memory.
The average speedup is 14\%, while individual workloads like $BigLSTM$ and
$VGG16$ achieve high speedups of 28\% and 30\%, respectively.  The reason for the
higher speedup in these workloads follows from Figures~\ref{fig:DL_footprint}
and ~\ref{fig:DL_sweep}.  Without compression, both of these are unable to fit
the mini-batch size of 64, which needed for good resource
utilization.

This overall speedup of $14\%$ is much higher than the $2.2\%$
performance overhead due to
Buddy Compression (Figure~\ref{fig:perf}). This indicates that Buddy
compression can lead to significant performance gain for capacity-constrained
GPUs by allowing the use of larger mini-batch sizes.

\textbf{Better Convergence with Larger Mini-Batches.}\label{sec:acc}
Apart from improving computational throughput with better resource utilization,
the mini-batch size can also impact the training accuracy.  In order to
investigate this, we train ResNet50 on the CIFAR100~\cite{cifar100} dataset for 100
epochs on a Titan Xp GPU with different mini-batch sizes.
Figure~\ref{fig:DL_train} shows the validation accuracy results for these runs.
As can be seen, very small mini-batches of 16 and 32 do not reach the maximum
accuracy, despite using the corresponding, tuned hyperparameters. Additionally,
although the mini-batch size of 64 trains to the maximum accuracy, it converges
slower than the larger mini-batches. With batch normalization, the jitter in
the accuracy is also higher with small mini-batch sizes.  While we observe good
validation accuracy up to a batch size of 256, which is in line with the reported
results in previous work~\cite{megdet2017}, it has been reported that
increasing the mini-batch beyond a certain size can be detrimental to the
network's generalization. However, there has been other work on tuning loss functions
and hyperparameters for successful training with large mini-batches~\cite{Imagenet_1hr,Effects_DataPar}.

\textbf{Huge DL Networks.} Recent object detection networks like MegDet~\cite{megdet2017} and natural language processing networks
like BERT~\cite{BERT} are unable to fit more than 2-4 input samples per GPU during training,
due to memory capacity limits.
This is a hurdle for developers, since the best regularization technique, Batch Normalization requires a
batch size of at least 32 samples to be effective~\cite{groupnorm}.
As a result, developers resort to horizontal scaling by spreading a mini-batch
across many GPUs.
As an example, the version of MegDet that won the COCO challenge~\cite{coco} in 2017, performs
batch normalization across 128 GPUs, resulting in high communication overhead.
They also present results proving that larger mini-batches lead to higher accuracy, and are faster to train.
Using horizontal scaling alone to support larger batches is not sustainable due to the
inter-GPU communication bottleneck.
While our simulation infrastructure is unable to support such huge DL training networks,
Buddy Compression enables modest vertical scaling, which, when combined with horizontal scaling can lead to
more sustainable solutions.

The final takeaway from this case-study is that most DL networks require
a mini-batch of at least 64 or 128 in order to achieve near-maximum throughput
and best accuracy (with batch normalization).
Buddy Compression can help achieve the required mini-batch
sizes for large networks using fewer GPUs.

%

\section{Conclusions} This work proposes and evaluates Buddy Compression: the first general-purpose mechanism that can be used to increase user visible GPU memory capacity on GPUs.
Buddy Compression is enabled by modern high-bandwidth interconnects that allow remote memory pool to be used as a backup when the compressibility is not sufficient.
Buddy Compression is able to achieve 1.5--1.9$\times$ memory compression ratios across a wide range of HPC and deep learning workloads while incurring only a 1--2\% performance penalty compared to a system with a larger GPU memory capacity, due to its unique design where compressibility changes do not incur additional data movement.
This combination of high performance and reasonable compression ratios makes Buddy Compression an attractive and performant alternative to existing technologies like Unified Memory oversubscription.


{
\bibliographystyle{sty/IEEEtran}
\footnotesize
\bibliography{bib/compress,bib/dnn_training_scaling_out,bib/dnn_training_recomputation,bib/dnn_training_precision,bib/dnn_training_offloading,bib/gpu_snapshot,bib/gpu_background}
}

\end{document}